\documentclass[
twocolumn,
 amsmath,amssymb,
 aps,
pre,
showkeys
]{revtex4-2}

\usepackage{graphicx}
\usepackage{dcolumn}
\usepackage{bm}

\usepackage{easyReview}

\usepackage{textalpha}

\usepackage{float}

\usepackage{graphicx} 			

\usepackage{here}

\usepackage{natbib}

\usepackage{psfrag}

\usepackage{varioref}

\usepackage[unicode=true,
linktocpage,
linkbordercolor	={0.5 0.5 1},
citebordercolor	={0.5 1 0.5},
breaklinks		=true,
linkcolor		=blue,
colorlinks   	= true, 
urlcolor     	= black,
linkcolor    	= blue, 
citecolor   	= blue, 
]{hyperref}

\usepackage{booktabs}

\usepackage{bm}

\usepackage[cal=dutchcal,calscaled=1]{mathalpha}

\usepackage[markup=Highlight,color=yellow]{pdfcomment}

\usepackage{mathtools}

\usepackage{amssymb}

\usepackage{cancel}

\usepackage{wrapfig}

\usepackage{multirow}

\usepackage{nicefrac}

\usepackage{tipa}

\usepackage{sidecap} 		

\usepackage[normalem]{ulem} 	

\usepackage{listings}
\lstdefinestyle{mystyle}{
	language=C++,
	frame=tblr,
	backgroundcolor=\color{backcolour},   
	commentstyle=\color{codegreen},
	keywordstyle=\color{magenta},
	numberstyle=\tiny\color{codegray},
	stringstyle=\color{codepurple},
	basicstyle=\ttfamily\tiny,
	breakatwhitespace=false,         
	breaklines=true,                 
	captionpos=t,                    
	keepspaces=true,                 
	numbers=left,                    
	numbersep=5pt,                  
	showspaces=false,                
	showstringspaces=false,
	showtabs=false,                  
	tabsize=2
}

\definecolor{codegreen}{rgb}{0,0.6,0}
\definecolor{codegray}{rgb}{0.5,0.5,0.5}
\definecolor{codepurple}{rgb}{0.58,0,0.82}
\definecolor{backcolour}{rgb}{0.95,0.95,0.92}

\definecolor{myorange}{RGB}{218, 84, 26}
\definecolor{myblue}{RGB}{0, 112, 192}
\definecolor{myyellow}{RGB}{237, 177, 32}
\definecolor{mypurple}{RGB}{126, 47, 142}
\definecolor{mygreen}{RGB}{119, 172, 48}
\definecolor{mycyan}{RGB}{77, 190, 238}
\definecolor{mydarkred}{RGB}{162, 20, 47}

\usepackage[linesnumbered,lined,boxed,commentsnumbered]{algorithm2e} 

\usepackage{verbatim}

\usepackage{fancyvrb}

\usepackage{wrapfig}

\usepackage{tabularx}

\usepackage{lipsum}

\usepackage[font=footnotesize,labelfont=bf]{caption}
\usepackage{subcaption}

\usepackage{inputenc}

\newcommand{\DKM}{\mathcal{D}^{(n)}}

\newcommand{\DKMv}{\mathcal{D}^{(1)}}
\newcommand{\DKMvv}{\mathcal{D}^{(2)}}

\newcommand{\Hbul}{\mathcal{H}_{\underline{\beta}}^{(m)}}
\newcommand{\Haul}{\mathcal{H}_{\underline{\alpha}}^{(l)}}
\newcommand{\Hgul}{\mathcal{H}_{\underline{\gamma}}^{(m)}}
\newcommand{\Hbaul}{\mathcal{H}_{\beta\underline{\alpha}}^{(l+1)}}
\newcommand{\HhermitepowLpOne}{\mathcal{H}_{\beta\underline{\alpha}}^{(l+1)}}
\newcommand{\HhermiteZero}{\mathcal{H}^{(0)}}
\newcommand{\HhermiteOne}{\mathcal{H}_{\alpha}^{(1)}}
\newcommand{\HhermiteTwo}{\mathcal{H}_{\alpha\beta}^{(2)}}
\newcommand{\HhermiteThree}{\mathcal{H}_{\alpha\beta\gamma}^{(3)}}

\newcommand{\F}{\mathcal{F}}
\newcommand{\Faul}{\mathcal{F}_{\underline{\alpha}}^{(l)}}
\newcommand{\Fgul}{\mathcal{F}_{\underline{\gamma}}^{(l)}}
\newcommand{\Fzero}{\mathcal{F}^{(0)}}
\newcommand{\Faone}{\mathcal{F}_{\alpha}^{(1)}}
\newcommand{\Fbone}{\mathcal{F}_{\beta}^{(1)}}
\newcommand{\Fab}{\mathcal{F}_{\alpha\beta}^{(2)}}
\newcommand{\Fabg}{\mathcal{F}_{\alpha\beta\gamma}^{(3)}}

\newcommand{\aul}{\underline{\alpha}}

\newcommand{\baul}{\beta\underline{\alpha}}
\newcommand{\gul}{\underline{\gamma}}

\newcommand{\Gaul}{\mathcal{G}_{\underline{\alpha}}^{(l)}}
\newcommand{\Gzero}{\mathcal{G}^{(0)}}
\newcommand{\Gaone}{\mathcal{G}_{\alpha}^{(1)}}
\newcommand{\Gbone}{\mathcal{G}_{\beta}^{(1)}}
\newcommand{\Gab}{\mathcal{G}_{\alpha\beta}^{(2)}}

\newcommand{\Maul}{M^{(m)}_{\underline{\alpha}}}

\newcommand{\opHatFP}{\widehat{\Omega}^\textrm{(FP)}}
\newcommand{\opHatFPi}{\widehat{\Omega}_i^\textrm{(FP)}}
\newcommand{\opHatFPione}{\widehat{\Omega}_i^\textrm{(FP,1)}}

\newcommand{\opHatL}{\widehat{\Omega}^{(\eta)}}

\newcommand{\opFP}{\Omega^\textrm{(FP)}}

\newcommand{\opFPaul}{\Omega^{\textrm{(FP)}}_{\underline{\alpha}}}
\newcommand{\opFPgul}{\Omega^{\textrm{(FP)}}_{\underline{\gamma}}}
\newcommand{\opFPi}{\Omega^\textrm{(FP)}_i}
\newcommand{\opFPLi}{\Omega^{(\textrm{FP},\eta)}_i}
\newcommand{\opFPine}{\Omega^{(\textrm{FP},n_{\e})}_i}
\newcommand{\opFPizero}{\Omega^\textrm{(FP,0)}_i}
\newcommand{\opFPione}{\Omega^\textrm{(FP,1)}_i}
\newcommand{\opFPitwo}{\Omega^\textrm{(FP,2)}_i}
\newcommand{\opBGK}{\Omega^\textrm{(B)}}
\newcommand{\opBGKi}{\Omega^\textrm{(B)}_i}

\newcommand{\opLaul}{\Omega^{(\eta)}_{\underline{\alpha}}}
\newcommand{\opLgul}{\Omega^{(\eta)}_{\underline{\gamma}}}
\newcommand{\opLi}{\Omega^{(\eta)}_i}
\newcommand{\opS}{\Omega^{(\textrm{S})}}

\newcommand{\omgBGK}{\omega^{\textrm{(B)}}}
\newcommand{\omgFP}{\omega^\textrm{(FP)}}

\newcommand{\E}{\mathcal{E}}
\newcommand{\Eneq}{\mathcal{E}^\textrm{(neq)}}
\newcommand{\Eone}{\mathcal{E}^{(1)}}
\newcommand{\Etwo}{\mathcal{E}^{(2)}}
\newcommand{\Eol}{\overline{\mathcal{E}}}

\newcommand{\Eolone}{\overline{\mathcal{E}}^{(1)}}
\newcommand{\Eoltwo}{\overline{\mathcal{E}}^{(2)}}
\newcommand{\bmx}{\bm{x}}

\newcommand{\bmk}{\bm{k}}
\newcommand{\bmkhat}{\widehat{\bm{k}}}
\newcommand{\dt}{\delta t}

\newcommand{\dab}{\delta_{\alpha\beta}}
\newcommand{\dad}{\delta_{\alpha\delta}}
\newcommand{\dbg}{\delta_{\beta\gamma}}
\newcommand{\dbd}{\delta_{\beta\delta}}
\newcommand{\dde}{\delta_{\delta\varepsilon}}
\newcommand{\ddz}{\delta_{\delta\zeta}}
\newcommand{\dez}{\delta_{\varepsilon\zeta}}
\newcommand{\dgd}{\delta_{\gamma\delta}}

\newcommand{\dlm}{\delta_{lm}}
\newcommand{\deltaag}{\delta_{\alpha\gamma}}
\newcommand{\daulbul}{\delta_{\underline{\alpha}\underline{\beta}}^{(l)}}
\newcommand{\dgulaul}{\delta_{\underline{\gamma}\underline{\alpha}}^{(l)}}

\newcommand{\vaeta}{v_\alpha^{(\eta)}}

\newcommand{\Al}{A_\parallel}
\newcommand{\ulong}{u_\parallel}

\newcommand{\bmut}{\bmu_\perp}

\newcommand{\vbeta}{v_\beta^{(\eta)}}

\newcommand{\ua}{u_\alpha}
\newcommand{\ub}{u_\beta}

\newcommand{\ug}{u_\gamma}
\newcommand{\ud}{u_\delta}
\newcommand{\ue}{u_\varepsilon}
\newcommand{\uz}{u_\zeta}
\newcommand{\vgeta}{v_\gamma^{(\eta)}}

\newcommand{\vzeta}{v_\zeta^{(\eta)}}
\newcommand{\bmveta}{\bm{v}^{(\eta)}}

\newcommand{\bmu}{\bm{u}}
\newcommand{\ja}{j_\alpha}
\newcommand{\jaone}{j_\alpha^{(1)}}
\newcommand{\jatwo}{j_\alpha^{(2)}}
\newcommand{\jaeta}{j_\alpha^{(\eta)}}

\newcommand{\jdneq}{j_\delta^\textrm{(neq)}}

\newcommand{\jdeq}{j_\delta^\textrm{(eq)}}
\newcommand{\jaol}{\overline{\jmath}_\alpha}

\newcommand{\Jaolen}{\overline{\jmath}_\alpha^{(n_{\e})}}
\newcommand{\jaolzero}{\overline{\jmath}_\alpha^{(0)}}
\newcommand{\Jaolone}{\overline{\jmath}_\alpha^{(1)}}
\newcommand{\Jaoltwo}{\overline{\jmath}_\alpha^{(2)}}
\newcommand{\Jb}{j_\beta}

\newcommand{\Jg}{j_\gamma}
\newcommand{\qa}{q_\alpha}

\newcommand{\qdneq}{q_\delta^\textrm{(neq)}}
\newcommand{\qdeq}{q_\delta^\textrm{(eq)}}

\newcommand{\qdone}{q_\delta^{(1)}}
\newcommand{\qatwo}{q_\alpha^{(2)}}
\newcommand{\qaol}{\overline{q}_\alpha}

\newcommand{\qaoltwo}{\overline{q}_\alpha^{(2)}}
\newcommand{\qb}{q_\beta}
\newcommand{\qd}{q_\delta}

\newcommand{\sqrtTta}{\sqrt{\theta}}
\newcommand{\realR}{\mathbb{R}}
\newcommand{\mathO}{\mathcal{O}}

\newcommand{\mathF}{\mathcal{F}}

\newcommand{\rhotilde}{\widetilde{\rho}}

\newcommand{\rhool}{\overline{\rho}}
\newcommand{\rhoolen}{\overline{\rho}^{(n_{\e})}}
\newcommand{\rhoolzero}{\overline{\rho}^{(0)}}
\newcommand{\rhoolone}{\overline{\rho}^{(1)}}
\newcommand{\rhooltwo}{\overline{\rho}^{(2)}}

\newcommand{\rhoneq}{\rho^\textrm{(neq)}}

\newcommand{\drho}{\delta\rho}
\newcommand{\Etld}{\widetilde{E}}
\newcommand{\ptilde}{\widetilde{p}}

\newcommand{\bmutilde}{\widetilde{\bm{u}}}

\newcommand{\cv}{c_v}
\newcommand{\specificcp}{c_p}

\newcommand{\delp}{\delta p}
\newcommand{\e}{\epsilon}
\newcommand{\bmeta}{\bm{\eta}}
\newcommand{\etaa}{\eta_{\alpha}}
\newcommand{\etab}{\eta_{\beta}}
\newcommand{\etag}{\eta_{\gamma}}

\newcommand{\etaz}{\eta_{\zeta}}

\newcommand{\muB}{\mu^\textrm{(B)}}
\newcommand{\muFPBpls}{\mu^\textrm{(FPB)}_{+}}
\newcommand{\muFPBmns}{\mu^\textrm{(FPB)}_{-}}
\newcommand{\nul}{\nu_\parallel}
\newcommand{\nut}{\nu_\perp}
\newcommand{\zB}{\zeta^\textrm{(B)}}
\newcommand{\zFPBpls}{\zeta^\textrm{(FPB)}_+}
\newcommand{\zFPBmns}{\zeta^\textrm{(FPB)}_-}

\newcommand{\parta}{\partial_\alpha}
\newcommand{\partaone}{\partial_\alpha^{(1)}}
\newcommand{\partdone}{\partial_\delta^{(1)}}
\newcommand{\parteone}{\partial_\varepsilon^{(1)}}
\newcommand{\partzone}{\partial_\zeta^{(1)}}

\newcommand{\partd}{\partial_\delta}
\newcommand{\parte}{\partial_\varepsilon}
\newcommand{\partz}{\partial_\zeta}
\newcommand{\partbone}{\partial_\beta^{(1)}}

\newcommand{\partt}{\partial_t}

\newcommand{\parttone}{\partial_t^{(1)}}
\newcommand{\partttwo}{\partial_t^{(2)}}

\newcommand{\partva}{\partial_{v_\alpha}}
\newcommand{\partvb}{\partial_{v_\beta}}

\newcommand{\Dtuone}{D_t^{(1)}}
\newcommand{\Dtu}{D_t}
\newcommand{\tottbmvone}{D_t^{(\bmv,1)}}

\newcommand{\vi}{v_{i}}
\newcommand{\via}{v_{i,\alpha}}
\newcommand{\va}{v_\alpha}
\newcommand{\vb}{v_\beta}
\newcommand{\vg}{v_\gamma}

\newcommand{\vib}{v_{i,\beta}}

\newcommand{\vig}{v_{i,\gamma}}
\newcommand{\vid}{v_{i,\delta}}
\newcommand{\vie}{v_{i,\varepsilon}}
\newcommand{\viz}{v_{i,\zeta}}

\newcommand{\bmv}{\bm{v}}
\newcommand{\bmvi}{\bm{v}_i}
\newcommand{\vT}{v_T}
\newcommand{\vTsq}{v_T^2}
\newcommand{\cssq}{\varsigma^2}
\newcommand{\vTsql}{v_T^{2l}}

\newcommand{\bmci}{\bm{c}_i}

\newcommand{\wi}{w_{i}}
\newcommand{\ca}{c_\alpha}
\newcommand{\cb}{c_\beta}

\newcommand{\fizero}{f_i^{(0)}}
\newcommand{\gizero}{g_i^{(0)}}
\newcommand{\fione}{f_i^{(1)}}
\newcommand{\gione}{g_i^{(1)}}
\newcommand{\fitwo}{f_i^{(2)}}
\newcommand{\gitwo}{g_i^{(2)}}
\newcommand{\fiprm}{f_i'}

\newcommand{\fieq}{f_i\supereq}
\newcommand{\fineq}{f_i\superneq}

\newcommand{\gieq}{g_i\supereq}
\newcommand{\gineq}{g_i\superneq}
\newcommand{\sumQ}{\sum_{i=0}^{Q - 1}}
\newcommand{\sumi}{\sum_{i}}

\newcommand{\sumlK}{\sum_{l=0}^K}
\newcommand{\summK}{\sum_{m=0}^K}
\newcommand{\summlK}{\sum_{m=l=0}^K}
\newcommand{\sumlinf}{\sum_{l=0}^\infty}

\newcommand{\fTld}{\tilde{f}}

\newcommand{\fvxt}{f(\bmv; \bmx, t)}
\newcommand{\fvxit}{f_i(\bmvi; \bmx, t)}
\newcommand{\gvxt}{g(\bmv; \bmx, t)}

\newcommand{\Fz}{F_\zeta}
\newcommand{\Pab}{P_{\alpha\beta}}

\newcommand{\Pdeneq}{P_{\delta\varepsilon}^\textrm{(neq)}}
\newcommand{\Pabone}{P_{\alpha\beta}^{(1)}}
\newcommand{\Pdeone}{P_{\delta\varepsilon}^{(1)}}
\newcommand{\Pabtwo}{P_{\alpha\beta}^{(2)}}

\newcommand{\Pdeeq}{P_{\delta\varepsilon}^\textrm{(eq)}}

\newcommand{\Pabol}{\overline{P}_{\alpha\beta}}

\newcommand{\Pabolen}{\overline{P}_{\alpha\beta}^{(n_{\e})}}
\newcommand{\Pabolzero}{\overline{P}_{\alpha\beta}^{(0)}}
\newcommand{\Pabolone}{\overline{P}_{\alpha\beta}^{(1)}}
\newcommand{\Paboltwo}{\overline{P}_{\alpha\beta}^{(2)}}

\newcommand{\Rab}{R_{\alpha\beta}}

\newcommand{\Rdeneq}{R_{\delta\varepsilon}^\textrm{(neq)}}
\newcommand{\Rdeeq}{R_{\delta\varepsilon}^\textrm{(eq)}}
\newcommand{\Rabone}{R_{\alpha\beta}^{(1)}}
\newcommand{\Rabtwo}{R_{\alpha\beta}^{(2)}}
\newcommand{\Rabol}{\overline{R}_{\alpha\beta}}

\newcommand{\Rabolone}{\overline{R}_{\alpha\beta}^{(1)}}
\newcommand{\Raboltwo}{\overline{R}_{\alpha\beta}^{(2)}}

\newcommand{\Hherm}{\mathcal{H}}

\newcommand{\Qabg}{Q_{\alpha\beta\gamma}}

\newcommand{\Qdezeq}{Q_{\delta\varepsilon\zeta}\supereq}

\newcommand{\tautld}{\widetilde{\tau}}
\newcommand{\tautldl}{\widetilde{\tau}_\parallel}
\newcommand{\tautldt}{\widetilde{\tau}_\perp}

\newcommand{\bmtautldt}{\widetilde{\bm{\tau}}_\perp}

\newcommand{\chil}{\chi_\parallel}

\newcommand{\bmq}{\bm{q}}
\newcommand{\bmqtld}{\widetilde{\bm{q}}}

\newcommand{\xii}{\xi_i}

\newcommand{\tauab}{\tau\subab}
\newcommand{\taude}{\tau\subde}
\newcommand{\tauabtld}{\widetilde{\tau}\subab}
\newcommand{\taudetld}{\widetilde{\tau}\subde}

\newcommand{\supereq}{^{\textrm{(eq)}}}
\newcommand{\superneq}{^{\textrm{(neq)}}}

\newcommand{\superzero}{^{(0)}}
\newcommand{\superone}{^{(1)}}
\newcommand{\supertwo}{^{(2)}}

\newcommand{\subeq}{_{\textrm{(eq)}}}

\newcommand{\subab}{_{\alpha\beta}}
\newcommand{\subde}{_{\delta\varepsilon}}

\newcommand{\subdez}{_{\delta\varepsilon\zeta}}

\newcommand{\subabgd}{_{\alpha\beta\gamma\delta}}

\newcommand{\suba}{_{\alpha}}
\newcommand{\subaa}{_{\alpha\alpha}}
\newcommand{\subd}{_{\delta}}

\newcommand{\subz}{_{\zeta}}
\newcommand{\subzz}{_{\zeta\zeta}}

\newcommand{\derpTv}{\biggl(\frac{\partial p}{\partial T}\biggr)_v}

\newcommand{\derprT}{\biggl(\frac{\partial p}{\partial \rho}\biggr)_T}
\newcommand{\derpTr}{\biggl(\frac{\partial p}{\partial T}\biggr)_\rho}

\newcommand{\dervTp}{\biggl(\frac{\partial v}{\partial T}\biggr)_p}

\newcommand{\derprhos}{\biggl(\frac{\partial p}{\partial \rho}\biggr)_s}

\newcommand{\derprhoT}{\biggl(\frac{\partial p}{\partial \rho}\biggr)_T}

\newcommand{\dereTp}{\biggl(\frac{\partial e}{\partial T}\biggr)_p}
\newcommand{\derhTp}{\biggl(\frac{\partial h}{\partial T}\biggr)_p}

\begin{document}
	
\setreviewson
\setreviewsoff

\preprint{APS/123-QED}

\title[Towards an LFPBM]{Towards a lattice-Fokker-Planck-Boltzmann model\\ of thermal fluctuations in non-ideal fluids}

\author{K. J. Petersen}
	\affiliation{University of British Columbia, Okanagan, Canada}
	
\author{J. R. Brinkerhoff}%
 	\email{\add{joshua.brinkerhoff@ubc.ca}}
	\affiliation{University of British Columbia, Okanagan, Canada}

\date{\today}

\begin{abstract}
	Microscopic thermal fluctuations are known to affect the macroscopic and spatio-temporal evolution of a host of physical phenomena central to the study of biological systems, turbulence, and reactive mixtures, among others. In phase-changing fluids metastability and nucleation rates of embryos are known to be non-trivially affected by thermal noise stemming from molecules' random velocity fluctuations, which ultimately determine the long-term growth, morphology, and decay of macroscopic bubbles in cavitation and boiling. We herein present the mathematical groundwork for a lattice-based solution of the combined Fokker-Planck and Boltzmann equations that by proxy solve the stochastic Navier-Stokes-Fourier equations and a non-ideal, cubic van der Waals equation of state. We present the derivation of the kinetic lattice-Fokker-Planck-Boltzmann equations facilitated by Gauss-Hermite quadrature, and show by multi-scale asymptotic analysis that the non-equilibrium dynamics in velocity space inherent to the Fokker-Planck equation manifest as stresses. The resulting coarse-grained lattice-Fokker-Planck-Boltzmann method (LFPBM) is attractive as its dynamics are hypothesized to continually evolve thermal fluctuations introduced into the thermo-hydrodynamic variables by initial conditions in a manner that obeys the fundamental fluctuation-dissipation balances. Simulations will be showcased in future publications.  
\begin{description}
\item[Usage]
Our theoretical development is inspired by the lattice-Fokker-Planck development in [\citeauthor{Moroni2006a} Phys. Rev. E 73, 066707] and employs Particles-on-Demand for kinetic theory presented in [\citeauthor{Reyhanian2020} Phys. Rev. E 102, 020103(R)] to mitigate numerical instability associated with high compressibility and non-ideality. 
 
\end{description}
\end{abstract}


\maketitle

\section{Introduction}

\subsection{Fluctuations, nucleation, and phase transitions}

Nucleation is responsible for initiating \textit{homogeneous} as well as \textit{heterogeneous} phase transitions in fluid systems, and occurs in a myriad of natural and engineered processes. Phase transitions are commonly described in literature to be either pressure-driven along isotherms (\textit{cavitating}) or thermally-driven along isobars (\textit{boiling}) but in reality evolve in a blended manner simultaneously along both the $p$ and $T$ dimensions \cite{Petersen2023a}. Cavitation dominates in highly-advective flows with large pressure gradients causing local pressures to fall below the saturation pressure, whereas boiling occurs in \add{diabatic} flows where local temperatures exceed the saturation temperature. The relevance of either of the modes is exemplified by technological applications such as medical drug delivery and acoustic therapies facilitated by cavitation \cite{Coussios2008}, turbo-machinery erosion and efficiency loss from cavitation \cite{Leroux2005}, pool boiling in cryogenic spills \cite{Petersen2023a}, as well as natural occurrences in skeletal-joint cavitation \cite{Kawchuk2015}, cavitation in the xylem of drought-stricken trees \cite{Cochard2006}, cavitation as a predation tool in mantis shrimp \cite{Patek2004}, to mention a few exotic cases. 

Inherent to cavitation and boiling is the property that phase change may occur either homogeneously or heterogeneously, and that phase-transitioning liquids can be \textit{metastable}. In heterogeneous phase change, the presence of nucleation sites---e.g. on the surface of suspended solid particles and in suspended undissolved gases in cavitation, or in microscale crevices in the surface topology of a solid boiling substrate---can non-trivially alter the effective liquid tensile strength by an order of magnitude in comparison to the homogeneous counterpart occurring in pure fluids \cite{Menzl2016, Lisi2017, Guillemot2012}, thus increasing the probability of nucleation events occurring.

Commonly, phase transitions are thought to occur instantaneously when the pressure falls below the saturation state, and conversely when the temperature exceeds the saturation temperature, when they in reality are relaxed processes in time. This means that the thermodynamic state of a stretched or superheated liquid in between its binodal and spinodal loci may remain in its original (liquid) state for a finite period of time until it is sufficiently perturbed \cite{Carey2018, Magaletti2020} where the phase stability is fundamentally dictated by entropy generation and the entropy-extremum principle \cite{Callen1985}. Such perturbations can by exemplification be pressure discontinuities in shock waves propagating on the macroscales, and/or thermal noise below and around the microscales where large ensembles of molecules can produce fluctuations in the density and temperature of a fluid thence provoking a phase transition \cite{Gallo2022, Lohse2016}. 

In addition to the above complexity, phase transitions statistically evolve across scales; new thermodynamic phases originate from the molecular scale and develops into the macroscale \cite{Gallo2020, Menzl2016}, and the size distributions of impurities span the nm-mm scales. Consequently, the nuclei density (number of nucleation sites/m$^3$), and the rate at which vapor phases grow during phase change markedly impacts the thermo-hydrodynamics of many flows including the cloud-shedding dynamics in cavitating hydrofoils \cite{Venning2022}, or hypothetically the Taylor-Helmholtz instability and conjugate heat transfer in pool boiling \cite{Petersen2023a}.   

Capturing the heterogeneous and metastable nature of phase-transitioning fluids in numerical simulations is rather difficult, and in computational, macroscopic simulations of industrial relevance it is frequently assumed that phase transitions occur homogeneously and instantaneously, as found in our recent literature reviews \cite{Petersen2023a, Petersen2021}. Due to the span of length scales in heterogeneous phase transitions, resolving nucleation sites in macroscale simulations is computationally intractable with contemporary computational resources. Thus, a more viable approach to the problem is to involve sub-grid modeling of unresolved nuclei that can act as nucleation sites. In cavitation and boiling simulations such heterogeneity in the form of undissolved microbubbles and solid surfaces, rigorously coupled with the physics of phase transformations and capillary effects of surface imperfections, is yet to be fully represented in computational fluid dynamics. Nevertheless, there do exist strategies for introducing sub-grid nuclei into simulations. In cavitation, works such as \cite{Bryngelson2019} evolved sub-grid microbubbles with a Rayleigh-Plesset equation including phase-change mass transfer. In the case of boiling, nucleation-site density models for pool boiling \cite{Yazdani2016, Sato2018} have proven useful, but still rely on empirically informed activation models. Otherwise, the computational fluid dynamics community predominantly rely on lower-fidelity, empirical models for predicting latent heat transfer in cavitation and boiling simulations \cite{Petersen2023a}, such as those of \citet{Kunz2000}, \citet{Schnerr2001}, and \citet{Zwart2004}, which account for nucleation sites via a constant nucleation-site volume fraction, but not for non-equilibrium thermodynamic effects nor the impact of micro-scale thermal fluctuations on nucleation.

Recently, \citet{Menzl2016} showed that in homogeneously cavitating water, the classical nucleation theory of \citet{Volmer1926, Farkas1927, Becker1935} combined with a macroscopic, thermal-noise-augmented Rayleigh-Plesset equation was able to reproduce nucleation rates from molecular-dynamics simulations demonstrating the non-trivial diffusive effects of molecular fluctuations in phase transitions. The concept of including thermal-noise in macroscopic simulations is the foundation of fluctuating-hydrodynamics theory (FHT) \add{conceived by \citet{Landau1956a}. 
\\
\indent The authors introduced \textit{classical} fluctuations (i.e. with non-quantum oscillation frequencies $\omega\ll k_BT/\hbar$) into macroscopic observables of the hydrodynamic equations by a zero-mean Gaussian random ``outside'' stress tensor $\tauabtld$ in the Navier-Stokes equation, and an outside heat flow vector $\bmqtld$ in the conduction equation $\bmq =-k\nabla T + \bmqtld$, both being mutually uncorrelated. The cornerstone of the FHT are the spatio-temporally ensemble-averaged, two-point correlations,
\begin{align}
	\label{eq:FDT_tautld_naive}
	\bigl\langle\tautld(\bmx,t)\otimes\tautld(\bmx',t')\bigr\rangle &= A\subabgd\delta(\bmx-\bmx')\delta(t-t'),\\
        \label{eq:FDT_qtld_naive}
	\bigl\langle\bmqtld(\bmx,t)\otimes\bmqtld(\bmx',t')\bigr\rangle &= A\subab\delta(\bmx-\bmx')\delta(t-t'),
\end{align}
which are tightly correlated with the rate of change of total entropy \cite{Landau1980a} through the variance tensors $A\subabgd,A\subab$. The averages are seminal and still derived today with the fluctuation-dissipation theorem (FDT) in stochastic models as they inherently validate the consistency of modeled fluctuations.} In recent works by \citet{Magaletti2020, Gallo2018, Gallo2020, Gallo2022} the authors combined the theory with Navier-Stokes simulations of boiling and cavitation \add{using the finite-volume method. Alternatively to imposing the fluctuations on the macroscale their introduction into kinetic theories was explored starting in the 1990's, specifically in the fluctuating lattice-Boltzmann method (FLBM) \cite{Succi2018} which by virtue of its molecular roots can be extended to include the interaction forces in liquids that ultimately are responsible for thermal fluctuations.}  

\subsection{Fluctuating kinetic theories on the lattice}
FLBM research originated with the work by \citet{Ladd1993a} who proposed a fluctuating lattice-Boltzmann model for colloidal suspensions. The Brownian motion of the suspension was manifested by separately tracking the Newtonian dynamics of solid particles and incorporating the thermal fluctuations directly into the governing equations for an isothermal immersion liquid. Using the same FHT of \citet{Landau1959a}, \citet{Ladd1993a} added \textit{coarse-grained} fluctuations exclusively to the fluid-stress tensor for a single-population lattice-Boltzmann equation where fluctuating populations were designed so as to obey the FDT \eqref{eq:FDT_tautld_naive}. The computation of the fluctuating stresses were mediated by solving a discrete Langevin-type equation (we will detail the Langevin equation in \S\ref{sec:The FPE}).
\\
\indent \citet{Adhikari2005a} extended the work of \citeauthor{Ladd1993a} with noise satisfying a FDT directly at the lattice level such that the fluctuations in mass, momentum, and stresses were consistently mechanically equilibrated for all wave vectors $\bmk \equiv k\bigl[\cos(\theta), \sin(\theta)\bigr]^T$. The authors in addition to the hydrodynamic observables (referred to as modes) noted the existence of local, non-hydrodynamic ``ghost modes'' \cite{Benzi1990a,Vergassola1990a} which must be accounted for in the FDT and models with single as well as multi-relaxation time kernels to maintain hydrodynamic isotropy and Galilean invariance. If unaccounted for, these modes may drain thermal fluctuations from the hydrodynamic modes. In simulations the computational overhead associated with the ghost modes was reported to be $\sim 10\%$ in runtime. The noise in the mentioned degrees of freedom was imposed by a set of coupled Langevin equations evolving discretely in time to predict the fluctuating densities $\xii$ in their single-population FLB model. The model was numerically realized by computing the covariance matrix of $n - D - 1$ noises of $n$ macroscopic variables in $D$ spatial dimensions after which the noise terms were projected back to lattice space using eigenvectors connecting the continuum moments with the lattice-Boltzmann velocity space. As such, the noise is incorporated in a ``top-down'' manner. With their algorithm they recommended setting $k_B T\le 1/3000$ to achieve accurate incompressibility, which is informed by the low-Mach-number stability threshold $k_B T \ll \cssq = 1/3$ common for all classical LB models.
\\ 
\indent In an alternative effort, \citet{Dunweg2007a} returned to the lattice-gas roots of the LBM and derived a generalized lattice-gas model in which ensembles of population densities locally realize probability distributions rather than mass densities. Then, the collision rules for the model were described as Monte-Carlo processes satisfying a detailed balance \cite{Gardiner1997} for the transition probability between pre and post-collision states. Particularly, a linear relaxation model exactly solving a Langevin equation for each mode was adopted. Thus, achieving correct ``thermalization''---i.e. FDT consistency---for each degree of freedom requires modeling of the same number of random variables. 
\\ 
\indent \citet{Gross2010a} furthered the FLBM by extending the theory to represent non-ideality with a square-gradient free-energy functional $\int d\bmx \bigl(\phi(\rho) + \kappa/2 (\nabla\rho)^2\bigr)$ and a double-well Landau potential $\phi$. The authors thoroughly assessed the thermalization of all degrees of freedom (density, momentum, stresses, internal energy, heat flux, and ghost modes) by first establishing the FDTs for each, including the structure factor for density fluctuations and the covariance matrix being the variance in the FDT for the Langevin fluctuating sources. The FDTs were then central in determining the ``equilibration ratios'' $\bigl\langle| \widetilde{M}\suba(\bmk)|^2\bigr\rangle/G\subaa(\bmk)$ in Fourier space for the fluctuating component of each mode in the vector $\widetilde{M}\suba$ from the diagonal components of the correlation matrix $G\subab$. Moreover, the parameter space of the covariance matrix was shown to correlate with interface width and predict the stability of simulations of multi-phase flows. Lastly, capillary fluctuations in a liquid-vapor interface were simulated to showcase the application of the FLBM to multi-phase flows. 
\\
\indent With the previous works the FLBM saw concerted development from the initial introduction of noise in the momentum-flux of a single component, ``simple'' ideal gas \cite{Ladd1993a}, through the equilibration of all mechanical degrees of freedom including ghost modes \cite{Adhikari2005a}, further study of mechanical equilibration by assessing detailed balances of the stochastic ``collisions'' in FLBMs \cite{Dunweg2007a}, and then extension of the FLBM to non-ideal, square-gradient fluids with focus on the full thermalization of mechanical as well as thermal modes \cite{Gross2010a}. \citet{Belardinelli2015a} generalized the work of \citet{Gross2010a} to address multi-component fluids.  A single FLBE was employed and the fluctuating strength of the various thermo-hydrodynamic modes were controlled by a multi-relaxation time model. The long-wavelength consistency in terms of FDTs for the degrees of freedom was analyzed with the linearized non-equilibrium thermodynamics theory by \citeauthor{Onsager1953a} \cite{Onsager1953a,deGroot1962a}. Furthermore, the authors investigated the thermalization in homogeneous as well as heterogeneous cases and showcased proper relaxation via equilibration ratios computed from simulations. 
\\
\indent Recently, \citet{Lulli2024a} utilized a FLBM to study metastability in phase transitions. Phase separation was brought about with a pseudo-potential (Shan-Chen) interaction-force model perturbed directly by the stochasticity imposed in a single FLBE. In simulations with open-source code implemented on CPU/GPU architectures, \citet{Lulli2023a} confirmed equilibrium behavior in metastable states perturbed by small fluctuations in the long-wavelength limit. The density structure factor was also numerically verified to high precision. 
\\
\indent As suggested by \citet{Belardinelli2015a} and \citet{Moroni2006a}, it may be worthwhile employing the Fokker-Planck equation (FPE) as an alternative basis to the FLBM for studying thermal fluctuations in non-ideal fluids. Accordingly, we bring a brief introduction to the equation. The FPE by \citet{Fokker1914} and \citet{Planck1917} has been suggested to be a suitable model of liquids \cite{Moroni2006} compared to the popularly employed Boltzmann equation in lattice-Boltzmann methods, which arguably is a powerful tool in studying phase-changing flows \cite{Petersen2021}. This is owed to the FPE being directly derived from the Langevin equation, which models the velocity space of particles including their chaotic fluctuations that arise due to the continuous interaction between the electro-static potentials of neighboring molecules and other external forces. The FPE thereby \textit{mimics} the stochastic behavior of the Langevin equation with an evolving probability distribution, from the underlying molecular dynamics. We are interested in investigating the suitability of the FPE as a versatile model of liquids; as the continuum density, momentum and total energy are recovered from the moments of the primitive variables of the FPE, the coarse-grained Langevin dynamics directly projects into the macroscopic variables as additional non-linear components $\widetilde{\bmu}, \widetilde{\rho}, \widetilde{p}$, etc. Moreover, entropy generated by thermal noise can be directly derived from the FPE \cite{Lucia2014} and as entropy maxima dictate phase stability \cite{Callen1985}, the FPE framework may present an avenue for rigorously studying the non-equilibrium residence times of fluids in metastable regimes. The infinitesimal oscillations in pressure and density fields are presumed to enable the destabilization of (respectively) stretched and superheated metastable fluids, and thereby spontaneous nucleation.

The FPE has already enjoyed success in investigating the turbulence-energy cascade and its associated entropy generation \cite{Peinke2019}, modeling rarefied-gas dynamics \cite{Gorji2011, Gorji2013, Gorji2014, Gorji2015}, ion transport in clays \cite{Rotenberg2006b}, homogeneous and heterogeneous condensation \cite{Ruckenstein2015}, among others. Thus, in a series of articles, we wish to assess the potential numerical and theoretical capabilities of the FPE tailored to nucleation and first-order phase transitions. As such, with this article we initially seek a lattice-based solution framework of the FPE for a simple, pure van der Waals fluid based on a cubic equation of state (EOS). We acknowledge that real noise in fluids may not be white, but attributed with a colored spectrum, pending further investigation in a dedicated literature review. Nevertheless, prior to exploring colored noise, we first wish to establish the lattice solution with idealized white noise. Eventually the work can be extended to accommodate more unique spectra for various fluids, more sophisticated EOS, as well as multi-component kinetic models. To understand how nucleation is mediated by thermal noise, section \S\ref{sec:The FPE} lays out the Langevin foundation of the FPE and explores its benefits as a kinetic model of liquids \cite{Succi2018}. 

\subsection{The Fokker-Planck equation}
\label{sec:The FPE}
First-order phase transitions occur when a critically large population of molecules experience sudden large fluctuations in the momenta that manifest a perturbation in entropy and hence a transition between stable states associated with the concavity of entropy \cite{Callen1985}. Since the 1970's the Fokker-Planck equation has proven a powerful tool for studying problems that involve stochastic fluctuations (noise), and how fluctuations affect systems around critical transition points \cite{Risken1984}. In describing the FPE, we start with the 1D linear Langevin equation for Brownian acceleration of a particle with mass $m$ and speed $\xi$ as,
\begin{align}
	\label{eq:Langevin}
	\eta = \partt \xi = - \gamma \xi + \widetilde{\Gamma}(t),
\end{align}
in which $\gamma \doteq \alpha/m = 1/\tau$ is a friction factor that correlates with a Stokes-like damping-force coefficient $\alpha$, and a relaxation time $\tau$ \cite{Risken1984}. For simplicity we only included the Langevin equation for a single velocity component, but it generalizes to a molecular-velocity vector $\bmv = \bigl[\xi_x, \xi_y, \xi_z\bigr]^\dagger$ in three dimensions. In the article we use bold-font for vector quantities \add{but regular font for higher-rank tensors including the viscous stresses $\tauab$}. The stochasticity in the equation is facilitated by the per-unit mass Langevin fluctuation force term $\widetilde{\Gamma} = \widetilde{F}/m$. Assuming that $\widetilde{\Gamma}(t)$ is Gaussian and $\delta$-correlated, a set of key properties arise from the fluctuation-dissipation theorem (FDT); namely that the average over an ensemble is zero $\bigl\langle \widetilde{\Gamma}(t)\bigr\rangle = 0$, and that the ensemble-average of a product of two Langevin force-terms at different times is zero if $t - t'$ is greater than the correlation time $\tau_c$, i.e. $\bigl\langle \widetilde{\Gamma}(t)\widetilde{\Gamma}(t')\bigr\rangle = 0, |t - t'| \ge \tau_c$. Conversely, when we are interested in timescales smaller than the correlation time, the ensemble average is Dirac $\delta$-correlated, 
\begin{align}
	\label{eq:FDT_Langevin}
	\bigl\langle \widetilde{\Gamma}(t)\widetilde{\Gamma}(t') \bigr\rangle = \add{A} \delta (t-t'),
\end{align}
where the noise-strength \add{(i.e. variance)} can be shown to be $A = 2\gamma k_B T/m$, where $k_B$ is Boltzmann's constant and $T$ is the absolute temperature. In a three dimensional case $\widetilde{\Gamma}(t)$ would be a vector quantity with mutually independent components for each of the spatial dimensions. Conversely, if the noise components were mutually dependent the strength $A$ would be a matrix quantity. The noise is ``white'' characterized by the spectral density of the Langevin equation \eqref{eq:Langevin} being independent of frequency. On the other hand, for non-$\delta$-correlated terms, the spectral density is frequency-dependent and hence the noise is attributed as being ``colored''. We here clarify that we are not interested in directly solving the Langevin equation for a finite number of particles, but instead wish to solve an equivalent equation of motion for a distribution $\fvxt$ of Langevin particles. With this notation of the arguments that we generally adopt in the article, we imply that $f$ is inherently dependent on the molecular velocities being the quadrature nodes of the population, whereas the spatio-temporal dependence is indirectly owed to the time-integration of the FPE, as well will detail later. The bridge between the FP and Langevin equations is the Kramers-Moyal (KM) expansion \cite{Kramers1940, Moyal1949} around a single or larger set of fluctuating variables, where we specifically consider the molecular velocity vector $\bmv$ in one and two spatial dimensions. If the Langevin equation with Gaussian $\delta$-correlated noise governs $\bmv$ it can be shown \cite{Risken1984} that the KM expansion,
\begin{align}
	\label{eq:KM_series}
	\partt f\bigl(\bmv;\bmx,t\bigr) = \sum_{n = 1}^{\infty} \bigl(-\partial_{\bmv}\bigr)^n \DKM f\bigl(\bmv;\bmx,t\bigr), 
\end{align}
reduces to the FPE, with vanishing coefficients $\DKM$ for the higher-order terms with $n \ge 3$. In this case the expansion comprises the deterministic drift coefficient $\DKMv\ (n = 1)$, and the diffusion coefficients $\DKMvv\ (n = 2)$ accounting for the fluctuations in the variable. Truncation to any finite order $n > 2$ results in introducing vanishingly small noise contributions from the Langevin equation, guaranteeing that the process is statistically continuous except for some non-vanishing cases \cite{Friedrich2011} where discrete, anomalous events play a significant role \cite{Peinke2019}. The drift and diffusion terms read,
\begin{gather}
	\label{eq:Langevin_drift_term}
	\DKMv 		= - \gamma\bmv -\bmeta,\\
	\label{eq:Langevin_diff_term}
	\DKMvv 		= 1/2q = \gamma k_B T/m, 
\end{gather}
which in concert with \eqref{eq:KM_series} for $\fvxt$ in position and velocity-space produces the FP equation \cite{Risken1984}, 
\begin{align}
	\label{eq:FPE_cont}
	\bigl(\partial_t + \va \parta + \etaa\partva\bigr)f = \opHatFP\circ f \doteq \gamma \partva\bigl(\va + \vTsq\partva\bigr)f, 
\end{align}
where $\vT = \sqrt{k_B Tm^{-1}}$ is the thermal velocity. We further appended the LHS body-forcing term with acceleration $\bm{F}_B/m = \bmeta$, which is not to be confused with the Langevin acceleration. We employ the notation for the FP operator $\opHatFP\circ f$ imposed on the populations, and apply the Einstein notation for which repeated subscripts denote summation over pertinent indices. At this point, we emphasize that the FP operator effectively models liquid states \cite{Moroni2006, Succi2018} and so we also need to consider a hard-sphere binary collision model for gaseous states. To that end, the Boltzmann operator $\opBGK \doteq \tau_c^{-1}\bigl(f\supereq - f\bigr)$ with the Maxwell-Boltzmann (MB) equilibrium distribution, 
\begin{align}
	\label{eq:feq}
	f\supereq 	= \frac{\rho}{\bigl(2\pi RT\bigr)^{D/2}}\exp\left(-\frac{\bigl(\bmv - \bmu\bigr)^2}{2RT}\right),
\end{align}
introduced into the RHS of \eqref{eq:FPE_cont} yields a Fokker-Planck-Boltzmann (FPB) type equation,
\begin{align}
	\label{eq:FPBE_cont}
	\bigl(\partial_t + \va \parta\bigr)f = 	-\etaa\partva f &+ \gamma \partva(\va + \vTsq\partva)f \nonumber\\
	&+ \frac{1}{\tau_c}\bigl(f\supereq - f\bigr), 
\end{align}
where the FP and Boltzmann operators share the same MB equilibrium \cite{Moroni2006}, and the acceleration term has been moved from the advection side. In the equilibrium \eqref{eq:feq} $\rho$ is the mass density, $R$ the gas constant, $D$ the number of spatial dimensions, $\bmv$ the molecular velocity, and $\bmu$ the continuum velocity. We are interested in achieving a unique, transient solution of the FPB equation in seven \add{discrete} degrees of freedom, i.e. $(\bmv,\bmx,t)$, and its moment equations enabling simulation of large-scale fluid-dynamics problems with phase transitions. The FP operator is discretized on the lattice following \citet{Moroni2006a} to derive kinetic equations for phase-changing flows of pure van der Waals fluids. The Boltzmann operator is modeled following the conventional lattice-Boltzmann method with the Bhatnagar-Gross-Krook (BGK) single-relaxation time collision model \cite{Bhatnagar1954a}. 

We emphasize here that in contrast to the aforementioned FLBM literature in which energy conservation was predicted by a square-gradient functional of $f$, we employ two populations $f$ and $g$ and two kinetic lattice FPBEs to separately evolve the hydrodynamics and thermodynamics. The FPE effectively models ``glancing'' interactions apt for representing liquids whereas the BGK operator purports to model collisions in gases \cite{Bhatnagar1954a}. It follows, that across a liquid-vapor interface, relaxation towards equilibrium is going to be mediated by a combination of glancing and hard-sphere collisions. A simple numerical realization of the two in combination is a superposition in which the operators' relative relaxation towards equilibrium is tuned by two independent relaxation frequencies. The approach is analogous to the formalism of the \citet{Skinner1980} collision kernel where the proportionality of the FP and BGK operators is captured in a single relaxation frequency. However, projecting the kernel onto the lattice is arguably more complex than treating the operators separately as in the current work. 

Moreover, it is understood that the Fokker-Planck equation evolves a deterministic distribution function $f$ not dissimilar to the Boltzmann formalism, albeit that the collision dynamics are markedly different. The FPE to an extent averages over the fluctuating Langevin states while retaining coarse-grained information about the fluctuating system at an arbitrary, characteristic long-wavelength limit overlapping with the continuum. Thus, the solution philosophy for \eqref{eq:FPE_cont} comprises evolving randomly-initialized macroscopic fields $\{\rho,p,\bmu\}(\bmx,t_0)$ in space, where the collision kernel continuously attempts to relax non-equilibrium states towards the MB equilibrium \eqref{eq:feq} coinciding with the stationary solution $\opFP\bigl(f\supereq\bigr) = 0$ satisfying detailed balance \cite{Reichl1998}. As relatively few computational realizations of \eqref{eq:FPE_cont} exist, its dynamical behavior is generally unclear, particularly in increasingly complex formulations including non-ideal equations of state, interaction potentials, capillarity, external forces, variable friction $\gamma(T)$, and so forth. It is the dynamics we wish to explore, with the aim of determining the extent to which granular thermal fluctuations are retained between the Langevin and FP levels, and if such fluctuations can be modulated with the thermo-hydrodynamics whilst obeying the fundamental fluctuation-dissipation balances including (\ref{eq:FDT_tautld_naive},\ref{eq:FDT_qtld_naive}) and analogous correlations for the remaining macroscopic variables. 

This paper revolves around the derivation of lattice solution of the FP operator, and is organized as follows: In \S\ref{sec:Kinetic equations} we seek a phase-space discretized form of \eqref{eq:FPBE_cont} via a Hermite-series expansion resulting in a set of explicit kinetic lattice equations with a discretized velocity space in accordance with conventional $DmQn$ lattice models, exemplified by the $D2Q9$ variant in particular. Then, in \S\ref{sec:CEexpansion} a Chapman-Enskog multiscale analysis is carried out to prove that the continuum conservation laws (the Navier-Stokes equations via $f$ and the energy equation via $g$) are satisfied by the lattice equations. A discussion in \S\ref{sec:Discussion} revolving around thermodynamic consistency, as well as strengths and weaknesses of the model, is followed by a conclusion \S\ref{sec:conclusion} giving an outlook on planned applications.

\section{Kinetic equations}
\label{sec:Kinetic equations}

\subsection{Advection}
The FPBE \eqref{eq:FPBE_cont} comprises two processes that need to be treated separately in prospective simulations. Going forward we attribute relaxation due to the BGK operator with ``collisions'', and that due to the FP operator with ``perturbations''. The LHS advection terms represent the advection of populations prior to particle collisions and perturbations, and it is responsible for the spatio-temporal discretization of $f$ \add{and $g$} in $(\bmx,t)$-space. It is well known that the stochastic sample path of a Langevin particle can be predicted by integration of the (exemplary 1D) It\^{o} stochastic-differential equation (SDE), 
	\begin{align}
		d\xi(t) 	= \DKMv\bigl(\xi(t), t\bigr)dt + \sqrt{\DKMvv\bigl(t\bigr)}dW(t),
	\end{align}
	in which the first term prefixed by $\DKMv$ accounts for the deterministic effects of the Langevin equation and the second term prefixed by $\DKMvv$ the stochastic dynamics of the Langevin equation as prescribed by the Wiener process $W(t)$ with the difference $dW(t) = \widetilde{\Gamma}(t)dt$ \cite{Gardiner1997}. The equivalence between the It\^{o} formalism and the FPE is drift and diffusion coefficients that occur in the SDE and \eqref{eq:FPBE_cont}.

As the fluctuating macroscopic fields affected by non-ideality and non-equilibrium effects arising in the populations $f(\bmv;\bmx,t)$ are expected to evolve non-linearly in time and incur strong gradients in the flow, numerical stability and accuracy are critically important. Therefore, we adapt the lattice-discrete velocity space to the local thermo-hydrodynamics by means of the recently proposed Particles-on-Demand \add{(PonD)} method in the form \cite{Dorschner2018}, 
\begin{subequations}
	\label{eq:bmvi}
	\begin{align}
		\bmvi 	= \sqrtTta\bmci + \bmu,
	\end{align}
	such that the inertial reference frame is shifted by the continuum velocity $\bmu(\bmx,t)$ rendering $\bmci$ the constant relative velocities as given in the conventional LBM, and the discrete velocities are further scaled by the local thermodynamics through the reduced pressure $\theta(\bmx,t)$, that are predicted by the FPB populations. The subscript $i$ corresponds to each of the $n$ quadrature vectors of the pertinent $DmQn$ model. We adopt the real-gas formulation, 
	\begin{align}
		\theta 	= \frac{p}{\rho RT_L},
	\end{align}
	where $T_L$ is the lattice temperature of a corresponding isothermal, ideal-gas realization of the LFPBE, represented by a known constant for every $DmQn$ lattice model \cite{Kallikounis2021, Reyhanian2022}.
\end{subequations}
The recast velocities \eqref{eq:bmvi} enable simulation of high-Mach and thermal flows owing to the fact that the Hermite-expansion of the equilibrium populations shared by both the FP and Boltzmann equations become exact and velocity-error free \cite{Dorschner2018, Reyhanian2021a, Reyhanian2022},
\begin{gather}
	\label{eq:fieq}
	\fieq 	= \rho\wi,\\
	\label{eq:gieq}
	\gieq 	= \rho\wi\left(2e - D\frac{p}{\rho} + \vi^2\right),
\end{gather}
informed by $\rho(\bmx,t), D$, the quadrature weights $\wi$, internal energy $e(\bmx,t)$, and real-gas pressure $p(\bmx,t)$. $\vi^2 = \bmvi\cdot\bmvi$ is understood to be the dot product. Mitigating velocity errors in a prospective simulation with a lattice-FPBE is especially attractive as previous lattice-solutions \cite{Moroni2006} of the FPE have suffered from significantly reduced stability limits with increasing Reynolds numbers in comparison to conventional lattice-Boltzmann models.

The benefits of the scheme are contrasted by the complexity of the required solution methods for the considered lattice equations; in the conventional LBM the discrete velocity sets ensure exact streaming between lattice nodes, but this is not the case with \eqref{eq:bmvi} that is more likely to incur populations streaming from off-lattice locations to lattice nodes. Thus, establishing the populations at off-lattice locations demands a predictor-corrector scheme and a reconstruction procedure on the basis of the adjacent known on-lattice populations. \add{The numerical solution strategy for the current lattice equations would generally follow the methodology laid out in \cite{Dorschner2018,Reyhanian2021a} with the addition of the solution of the FP operator. The exact solution strategy will be detailed in a future publication}. Moreover, the off-lattice nature of the discrete velocities requires that spatio-temporal integration of \eqref{eq:FPBE_cont} is carried out from an instance $t-\dt$ back in time along the characteristic lines. We adopt the second-order accurate trapezoidal integration resulting in the kinetic lattice equations, 
\begin{subequations}
	\label{eq:LFPBEs}
	\begin{align}
		f_i(\bmx,t) &- f_i(\bmx - \bmvi\dt, t-\dt) 	\nonumber\\ 
		&= \omgFP\opFPLi(f_i) + \omgBGK\opBGKi(f_i), \\
		g_i(\bmx,t) &- g_i(\bmx - \bmvi\dt, t-\dt) \nonumber\\
		&= \omgFP\opFPLi(g_i) + \omgBGK\opBGKi(g_i),
	\end{align}
	where the populations $f$ and $g$ recovering density and total energy, respectively, are relaxed by the relaxation frequencies $\omgFP$ and $\omgBGK$ that are imposed on the lattice-discretized FP-perturbation $\opFPi$ and BGK-collision $\opBGKi$ operators.
\end{subequations}
In the temporal discretization of \eqref{eq:LFPBEs}, the streaming operator results in,
\begin{align}
	\int_{t-\dt}^{t}\opS\circ f(t)\ dt		&= f' - f\nonumber\\
	&= \frac{\dt}{2}\Bigl(\Omega\circ f' + \Omega\circ f\Bigr), 
\end{align}
where we adopt the notation for the advected $f' = f(\bmx,t)$, and the initial $f = f(\bmx - \bmv\dt, t - \dt)$ populations. However, the purpose of the integration is to be able to predict the advected state $f'$, and thus it is unfeasible to compute $\Omega\circ f'$ unless a transformation is made to avoid the implicitness of the equation. An explicit form can be procured analogously to the BGK model with the population remapping, 
\begin{align}
	\fTld 	\doteq f - \frac{\dt}{2}\Omega\circ f,
\end{align} 
which in the BGK case yields the kinetic equation \citep{Kruger2017}, 
\begin{align}
	\fTld' - \fTld 	&= -\omgBGK \opBGK\circ f \nonumber\\
	&= -\frac{\tau^{-1}\dt}{1 + \tau^{-1}\dt/2}\opBGK\circ \fTld.
\end{align}
It was previously shown \cite{Rotenberg2006a} that the trapezoidal integration of the FP operator yields a similar relaxation frequency, although instead mediated by the friction factor $\gamma$,
\begin{align}
	\label{eq:omgFP}
	\omgFP 		= \frac{\gamma\dt}{1 + \gamma\dt/2}.
\end{align}
In the BGK case, the zeroth-through-second moments coincide for both $f, \fTld$, but this is not the case for the FP operator in which only the zeroth moments coincide. As such, sampling to compute the higher-order moments $\sumi f_i\bmvi, \sumi f_i \via\vib$ must be done on the basis of $f_i$ (and similarly for $g_i$). This is relevant as the moments are necessary for carrying out the Hermite-series expansion, as we will present in the following.   

\subsection{Collision and perturbation}
Following \citet{Reyhanian2021a}, we adopt the exact equilibria in (\ref{eq:fieq},\ref{eq:gieq}) in the BGK collision model. However, we still need to establish a lattice model for the FP operator, which will be facilitated with a Hermite expansion. In the following we repeat some of the work already documented in the article by \citet{Moroni2006a}, but do so for the sake of completeness. However, as we will see later, our explicit formulations for the operator is different from that reported by \citeauthor{Moroni2006a} owing to the characteristic eigenvalue property of the \add{FP-operator} modified by the discrete velocity set \eqref{eq:bmvi} of PonD. We adopt a notation for the operators in which circumflex-accented (\ $\widehat{\phantom{a}}$\ ) operators denote that the operators are to be considered only as the operational expressions excluding the populations. This notation is important in the derivation of the expansions. The discretization of the operators as well as the populations $f,g$ is facilitated by a Hermite-polynomial tensor $\Haul(\bmv)$ basis in $D$ dimensions. Just as these tensors can be exploited to transform gradient terms in the continuous FPE \eqref{eq:FPE_cont}, the continuous population $\fvxt$ in phase-space can be expanded in terms of its Hermite coefficients $\Faul(\bmx,t)$ as \cite{Moroni2006a}
\begin{align}
	\label{eq:fxvt_HermExp_infty}
	\fvxt	= w(\bmv)\sumlinf \frac{1}{\vT^{2l}l!}\Faul\Haul(\bmv),
\end{align}
where the populations are proportional to the Gaussian weight function \cite{Moroni2006a}, 
\begin{align}
	\label{eq:GaussianWeight}
	w(\bmv) 	= \frac{1}{\bigl(2\pi\vTsq\bigr)^{D/2}}\exp\left[-\frac{(\bmv - \bmu)^2}{2\vTsq}\right].
\end{align}
The subscript $\underline{\alpha} = \alpha_1\cdots\alpha_l$ in the Hermite coefficients and polynomials contracts the $l$ indices of the $l$-rank Hermite tensor $\Haul$. To that end, the zeroth-through third-rank tensors are \citep{Grad1949a, Grad1949b}, 
\begin{subequations}
	\label{eq:HermitePoly}
	\begin{gather}
		\HhermiteZero(\bmv) 	= 1,\\
		\HhermiteOne(\bmv) 		= \va,\\
		\HhermiteTwo(\bmv) 		= \va\vb - \vT^2\dab,\\
		\HhermiteThree(\bmv) 	= \va\vb\vg - \vT^2\bigl(\va\dbg + \vb\deltaag + \vg\dab\bigr),
	\end{gather}
\end{subequations}
and further detailed by \citet{Moroni2006a}. The Hermite coefficients for the populations are the velocity integrals which occupy $\realR^D$,
\begin{align}
	\Faul(\bmx,t) 	= \int d\bmv~ \fvxt\Haul(\bmv).
\end{align}
To obtain a tenable series expansion, we truncate \eqref{eq:fxvt_HermExp_infty} at an order $K$ over $l$ ranks and discard those exceeding $K$, such that we approximate $\fvxt$ as \cite{Moroni2006a}, 
\begin{align}
	\label{eq:fxvt_HermExp_trunc}
	\fvxt 	= w(\bmv)\sumlK \frac{1}{\vT^{2l}l!}\Faul\Haul(\bmv). 
\end{align} 
By the Gauss-Hermite quadratures, we seek to carry out the velocity integrals at a finite set of velocity quadrature-nodes, or abscissae, such that the continuous integrals become $Q$ easily-digestible discrete sums over the index $i = 0, ..., Q-1$. Crucially, given any polynomial $p(\bmv)$ with a lower order than $2K$, the continuous integral of its product with a Gaussian function becomes the quadrature \cite{Moroni2006a}, 
\begin{align}
	\int d\bmv ~ w(\bmv)p(\bmv) 	= \sumQ w_i p(\bmvi),
\end{align}
accompanied by $\bmvi\in\realR^D$ and $w_i$ identified as the quadrature nodes \eqref{eq:bmvi} and weights, respectively. In \eqref{eq:fxvt_HermExp_trunc}, $f/w$ can be characterized as a maximum $K$-order polynomial, which conveniently can be exploited to define the familiar discrete populations $f_i,g_i$. Starting with the conventional $m$-order moments of $f$, 
\begin{align}
	\label{eq:Maul_f}
	\Maul \doteq \int d\bmv ~ \fvxt v_{\alpha_1} \cdots v_{\alpha_m},
\end{align} 
can be rewritten as \cite{Moroni2006a}, 
\begin{align}
	\int d\bmv ~ &\frac{w(\bmv)}{w(\bmv)}\fvxt v_{\alpha_1} \cdots v_{\alpha_m} \nonumber\\
	&= \sumQ \frac{w_i}{w(\bmvi)}\fvxit v_{i,\alpha_1} \cdots v_{i,\alpha_m} \nonumber\\
	&\doteq \sumQ f_i v_{i,\alpha_1} \cdots v_{i,\alpha_m},
\end{align} 
where we have arrived at the familiar discrete population $f_i$ by the ratio,  
\begin{align}
	\frac{f_i(\bmx,t)}{\fvxit} = \frac{w_i}{w(\bmvi)}.
\end{align}
$K \ge 2$ is required to reproduce the correct hydrodynamic behavior governed by the Navier-Stokes-Fourier (NSF) equations. The series up to these $K$-orders require the conventional quadratures of the $f_i,g_i$ populations which represent mass and total energy, respectively \cite{Kruger2017, Reyhanian2021a}, 
\begin{subequations}
	\label{eq:continuous_moments_f}
	\begin{gather}
		\label{eq:rho}
		\rho 	\doteq \int d\bmv~ \fvxt 				= \sumQ f_i,\\
		\label{eq:ja}
		\ja 	\doteq \int d\bmv~ \fvxt \va 			= \sumQ f_i\via,\\
		\Pab 	\doteq \int d\bmv~ \fvxt \va\vb 		= \sumQ f_i\via\vib,\\
		\Qabg 	\doteq \int d\bmv~ \fvxt \va\vb\vg 	= \sumQ f_i\via\vib\vig,
	\end{gather}
\end{subequations}
and
\begin{subequations}
	\label{eq:continuous_moments_g}
	\begin{gather}
		\label{eq:E}
		\E 			\doteq \int d\bmv~ \gvxt 				= \sumQ g_i,\\
		\qa 		\doteq \int d\bmv~ \gvxt \va 			= \sumQ g_i\via,\\
		\Rab 		\doteq \int d\bmv~ \gvxt \va\vb 		= \sumQ g_i\via\vib.
	\end{gather}
\end{subequations}
Here, we emphasize that $\E = 2\rho E$, and $\rho, \ua = \ja/\rho, E = e + \ua\ua/2$ are the macroscopic observables density, continuum velocity, and total energy, which are of the interest in prospective simulations, and recoverable as in any other conventional lattice-Boltzmann model. They also participate in the hydrodynamic limit derived in \S\ref{sec:CEexpansion}. Furthermore, we can observe that the Hermite coefficients $\Faul,\Gaul$ are merely linear combinations of these, and read, 
\begin{subequations}
	\label{eq:Faul}
	\begin{gather}
		\Fzero	 	= \rho, \\
		\Faone 		= \ja, \\
		\Fab 		= \Pab - \vTsq\rho\dab,\\
		\Fabg  		= \Qabg - \vTsq\bigl(\dab\Jg + \deltaag\Jb + \dbg\ja\bigr),  
	\end{gather}
\end{subequations}
for $f_i$ and, 
\begin{subequations}
	\label{eq:Gaul}
	\begin{gather}
		\Gzero 		= \E,\\
		\Gaone 		= \qa, \\
		\Gab 		= \Rab - \vTsq\bigl(2\rho E\bigr)\dab,
	\end{gather}
\end{subequations}
for $g_i$. In the following we will establish the Hermite series for $\opFPi$ and $\opLi$. As these reside in the exact same subspace as \eqref{eq:fxvt_HermExp_trunc}, we can utilize the quadrature of $\Faul$ and $\Gaul$ in the series.

\subsubsection{The lattice Fokker-Planck operator}
\label{sec:The lattice Fokker-Planck operator}
We are now prepared to apply the above methodology on the operators $\opHatFP = \gamma\partva\bigl(\va + \vTsq\partva\bigr)$ and $\opHatL = -\etaa\partva$ in order to derive explicit functionals for $\opFPi\bigl(f_i;g_i\bigr)$ and $\opLi\bigl(f_i;g_i\bigr)$. We start with the FP operator. Extending the function $\opHatFP\circ f$ by $w_i/w(\bmvi)$ enables us to write the action of the Hermite expansion on that function \cite{Moroni2006a}, 
\begin{subequations}
	\label{eq:Herm_series_opFP}
	\begin{align}
		\label{eq:Herm_series_opFP_a}
		\opFPi 		= w_i\sumlK\frac{1}{\vTsql l!} \opFPaul\Haul (\bmvi),
	\end{align}
	where the Hermite coefficient for the operator is computed as, 
	\begin{align}
		\opFPaul 	= \int d\bmv ~ \opHatFP\circ f\Haul(\bmv).
	\end{align}
\end{subequations}
We are interested in finding an operational expression that can be used to rewrite $\opHatFP$ in a more convenient, explicit form. The product $\opHatFP\bigl[w(\bmv)\Haul(\bmv)\bigr]$, which is identifiable in \eqref{eq:Herm_series_opFP} by insertion of \eqref{eq:fxvt_HermExp_trunc}, can be constructed and will comprise a set of gradient and divergence terms up to the second order. \add{To expand the product we need to introduce an important eigen-value property of the characteristic Gaussian $w(\bmv)$ which make the continuous-velocity partial derivatives directly solvable. In App. \ref{app:Eigenfunctions} we use the property to expand the product, where the former is found to be,}
\begin{align}
	\partvb w(\bmv) 	= -\frac{\sqrtTta\cb}{\vTsq}w(\bmv),
\end{align}
which indeed is an eigenfunction with the eigenvector $-\sqrtTta\cb/\vTsq$. This result accounts for the recast velocities \eqref{eq:bmvi}, whereas the corresponding eigenvalue $-\cb/\vTsq$ found in the analysis of \citet{Moroni2006a} considers the conventional peculiar velocity set $\bmci\in\realR^Q$ used by all LB models with exact streaming. As we demonstrate in App. \ref{app:Hermite_coefficients}, the operational expression derived from $\opHatFP\bigl[w(\bmv)\Haul(\bmv)\bigr]$ can be used to compute the Hermite coefficients $\opFPaul$, melded with the previously found Hermite coefficients for $\fvxt$ and $\Faul$. With the eigenvalue property, we find that the product expands to, 
\begin{align}
	\label{eq:opFP_w_H_operationalExpression}
	\opHatFP\bigl[w(\bmv)\Haul(\bmv)\bigr] = \gamma \left(1 + \ub\frac{\vib + \ub}{\vTsq} - l\right)\times\nonumber\\
	\bigl[w(\bmv)\Haul(\bmv)\bigr] - \gamma \frac{2\ub}{\vTsq}\bigl[w(\bmv)\Hbaul(\bmv)\bigr],
\end{align} 
whereas \citet{Moroni2006a} reported the result $\opHatFP\Bigl[w(\bmv)\Haul(\bmv)\Bigr] = -\gamma l \Bigl[w(\bmv) \Haul(\bmv)\Bigr]$ with the peculiar velocity set. In \eqref{eq:opFP_w_H_operationalExpression} $\vib$ are the $Q$ recast velocity vectors \eqref{eq:bmvi}, and $\Hbaul$ is the $(l+1)$--rank tensor relative to the current rank being considered for $\Haul$. Thus, the local thermo-hydrodynamics imposed by $\bmvi$ directly affect the FP dynamics. As \eqref{eq:opFP_w_H_operationalExpression} is fully explicit and $\Faul$ \eqref{eq:Faul} are merely linear combinations of the population moments \eqref{eq:Maul_f}, we have all the ingredients for computing the Hermite coefficients $\opFPaul$. The first three are computed in Tab. \ref{tab:Hermite_coeffs} using the explicit form of the coefficients \eqref{eq:opFPgul_app}. To compute the integral in the coefficients $\opHatFP$ we have utilized the Hermite orthonormality relation as is also showcased in App. \ref{app:Hermite_coefficients}. These coefficients, combined with \eqref{eq:Herm_series_opFP_a}, yield the series expanded up to $l \le K = 2$, for both $f_i,g_i$: 
\begin{widetext}
	\begin{subequations}
		\begin{align}
			\label{eq:opFPi_fi}
			\opFPi(f_i) 	&= w_i\gamma\Biggl[\rho\left(1 + \ug\frac{\vig + \ug}{\vTsq}\right) + \frac{\via}{\vTsq}\left(\ug\frac{\vig + \ug}{\vTsq} - 2\right)\ja \nonumber \\
			&+ \frac{\via\vib - \vTsq\dab}{2\vT^4}\left(\left\{\ug\frac{\vig + \ug}{\vTsq} - 1\right\}\left\{\Pab - \vTsq\rho\dab\right\} - 2\left\{\ua\Jb + \ub\ja\right\}\right) \Biggr],\\
			\label{eq:opFPi_gi}
			\opFPi(g_i) 	&= w_i\gamma\Biggl[\E\left(1 + \ug\frac{\vig + \ug}{\vTsq}\right) + \frac{\via}{\vTsq}\left(\ug\frac{\vig + \ug}{\vTsq} - 2\right)\qa \nonumber \\
			&+ \frac{\via\vib - \vTsq\dab}{2\vT^4}\left(\left\{\ug\frac{\vig + \ug}{\vTsq} - 1\right\}\left\{\Rab - \vTsq(2\rho E)\dab\right\} - 2\left\{\ua\qb + \ub\qa\right\}\right) \Biggr].
		\end{align}
	\end{subequations}
\end{widetext}
A similar analysis can be carried out for $\opHatL$ via its corresponding $\opHatL\Bigl[w(\bmv)\Haul(\bmv)\Bigr]$ function. 

\begin{table*}[]
	\caption{Hermite coefficients for the FP and acceleration lattice operators evaluated with (\ref{eq:opFPgul_app}, \ref{eq:opLgul_app}) as well as the Hermite coefficients for the populations $\Faul(f_i)$ \eqref{eq:Faul} and $\Gaul(g_i)$ \eqref{eq:Gaul}. The operator coefficients are ultimately used to complete the series expansion yielding $\opFPi$ and $\opLi$.}
	\label{tab:Hermite_coeffs}
	\begin{ruledtabular}
		\begin{tabular}{lll}
			$l\quad$ & $\opFPaul(l)\circ\{f_i;g_i\}$ & $\opLaul(l)\circ\{f_i;g_i\}$ \\
			\midrule
			0 & $\gamma\left[1 + \ug\frac{\vig + \ug}{\vTsq}\right]\Bigl\{\Fzero;\Gzero\Bigr\}$ 	& $-\left[\frac{\etag\ug}{\vTsq}\right]\Bigl\{\Fzero; \Gzero\Bigr\}$ \vspace{0.1cm}\\
			1 & $\gamma\left[\ug\frac{\vig + \ug}{\vTsq} - 2\right]\Bigl\{\Faone;\Gaone\Bigr\}$ 		& $\left[\etaa\Bigl\{\Fzero;\Gzero\Bigr\} - \frac{\etag\ug}{\vTsq}\right]\Bigl\{\Faone;\Gaone\Bigr\}$ \vspace{0.1cm}\\
			2 & $\gamma\Bigl[\left(\ug\frac{\vig + \ug}{\vTsq} - 1\right)\Bigl\{\Fab;\Gab\Bigr\} \quad$ & $\Bigl[\etaa\Bigl\{\Fbone;\Gbone\Bigr\} + \etab\Bigl\{\Faone;\Gaone\Bigr\} - \frac{\etag\ug}{\vTsq}\Bigl\{\Fab;\Gab\Bigr\}\Bigr]$ \vspace{0.1cm}\\
			& $\qquad - 2\Bigl(\ua\Bigl\{\Fbone;\Gbone\Bigr\} + \ub\Bigl\{\Faone;\Gaone\Bigr\}\Bigr)\Bigr]$ & 
		\end{tabular}
	\end{ruledtabular}
\end{table*}

\subsubsection{The lattice acceleration operator}
\label{sec:The lattice-Liouville operator}

\citet{Moroni2006a} found the operational expression $\opHatL\Bigl[w(\bmv)\Haul(\bmv)\Bigr] = \frac{\etab}{\vTsq}\Bigl[w(\bmv)\Hbaul(\bmv)\Bigr]$ with the peculiar velocity set. With the recast velocities, we instead report the result, 
\begin{align}
	\label{eq:opL_w_H_operationalExpression}
	\opHatL\Bigl[w(\bmv)\Haul(\bmv)\Bigr] 	= \frac{\etab}{\vTsq}\Bigl[w(\bmv)\bigl(\Hbaul(\bmv) - \ub\Haul(\bmv)\bigr)\Bigr].
\end{align}
This result can again be \add{expanded} in the same Hermite series \eqref{eq:Herm_series_opFP_a} as for the FP operator, recognizing that we are aiming at obtaining the combined lattice operator $\opFPLi = \opFPi + \opLi$, 
\begin{subequations}
	\label{eq:Herm_series_opL}
	\begin{align}
		\label{eq:Herm_series_opL_a}
		\opLi 		= w_i\sumlK\frac{1}{\vTsql l!} \opLaul\Haul (\bmvi),
	\end{align}
	where the Hermite coefficient for the operator is computed as, 
	\begin{align}
		\opLaul 	= \int d\bmv ~ \opHatL[f]\Haul(\bmv).
	\end{align}
\end{subequations}
Using \eqref{eq:opL_w_H_operationalExpression}, we get the relation for the Hermite coefficients \eqref{eq:opLgul_app} derived in App. \ref{app:Hermite_coefficients}, which we use to obtain the first three Hermite coefficients in Tab. \ref{tab:Hermite_coeffs}. These coefficients, in concert with \eqref{eq:Herm_series_opL_a} yield the full $(l \le K = 2)$--kinetic operators, 
\begin{subequations}
	\begin{align}
		\label{eq:opLi_fi}
		\opLi(f_i) 	&= w_i\gamma\Biggl[-\frac{\vgeta\ug}{\vTsq}\rho + \frac{\via}{\vTsq}\left(\vaeta\rho - \frac{\vgeta\ug}{\vTsq}\ja\right)\nonumber\\
		&\qquad + \frac{\via\vib - \vTsq\dab}{2\vT^4}\biggl(\left\{\vaeta\Jb + \vbeta\ja\right\} \nonumber\\
		&\qquad - \frac{\vgeta\ug}{\vTsq}\left\{\Pab - \vTsq\rho\dab\right\}\biggr)\Biggr],\\
		\label{eq:opLi_gi}
		\opLi(g_i) 	&= w_i\gamma\Biggl[-\frac{\vgeta\ug}{\vTsq}\E + \frac{\via}{\vTsq}\left(\vaeta\E - \frac{\vgeta\ug}{\vTsq}\qa\right)\nonumber\\
		&\qquad + \frac{\via\vib - \vTsq\dab}{2\vT^4}\biggl(\left\{\vaeta\qb + \vbeta\qa\right\} \nonumber\\
		&\qquad - \frac{\vgeta\ug}{\vTsq}\left\{\Rab - \vTsq\E\dab\right\}\biggr)\Biggr],
	\end{align}
\end{subequations}
where we adopted $\vaeta = \etaa\gamma^{-1}$ as a pseudo velocity designed by the dimensionality of $[\gamma] = s^{-1}$. Having rewritten the acceleration term this way, enables us to factorize terms with corresponding ranks of $\Faul,\Gaul$ into the same groups as all coefficients now are prefixed by $\gamma$. Having discretized both operators on the lattice we can combine them in more compact, convenient forms, 
\begin{subequations}
	\begin{align}
		\label{eq:opCombi_fi}
		\opFPLi(f_i) 	&= \opLi + \opFPi \nonumber\\
		&= w_i\gamma\left[\rhool + \frac{\via}{\vTsq}\jaol + \frac{\via\vib - \vTsq\dab}{2\vT^4}\Pabol\right],\\
		\label{eq:opCombi_gi}
		\opFPLi(g_i) 	&= \opLi + \opFPi \nonumber\\
		&= w_i\gamma\left[\Eol + \frac{\via}{\vTsq}\qaol + \frac{\via\vib - \vTsq\dab}{2\vT^4}\Rabol\right],
	\end{align}
	where we recasted the $f$-moments yielding density, momentum flux, and pressure and higher-order tensors into new functionals,
	\begin{align}
		\label{eq:moment_functionals_f_rhool}
		&\quad\rhool 	= \left(1 + \ug\frac{\vig + \ug - \vgeta}{\vTsq}\right)\rho, \\
		&\quad\jaol 	= \jaeta + \left(\ug\frac{\vig + \ug - \vgeta}{\vTsq} - 2\right)\ja, \\
		\label{eq:moment_functionals_f_Pabol}
		&\quad\Pabol	= \left(\vaeta - 2\ua\right)\Jb + \left(\vbeta - 2\ub\right)\ja \nonumber\\
		&\qquad + \left(\ug\frac{\vig + \ug - \vgeta}{\vTsq} - 1\right)\bigl(\Pab - \vTsq\rho\dab\bigr),
	\end{align}
	where $\jaeta = \rho\vaeta$ is a pseudo momentum. We have omitted the quadrature-node index notation in the recast moments and note that their dependence on $\vig$ is implied going forward. Evidently, $\vgeta$ originates from the redefinition of $\etag$. Similarly the recast functionals for the $g_i$-population moments read,
	\begin{align}
		&\quad\Eol 	= \left(1 + \ug\frac{\vig + \ug - \vgeta}{\vTsq}\right)\E, \\
		\label{eq:moment_functionals_g_qaol}
		&\quad\qaol 		= \ug\frac{\vig + \ug - \vgeta}{\vTsq}\qa + \Bigl(\vaeta - 2\ua\Bigr)\E, \\
		\label{eq:moment_functionals_g_Rabol}
		&\quad\Rabol		= \Bigl(\vaeta - 2\ua\Bigr)\qb + \Bigl(\vbeta - 2\ub\Bigr)\qa \nonumber\\
		&\qquad + \left(\ug\frac{\vig + \ug - \vgeta}{\vTsq} - 1\right)\Bigl(\Rab - \vTsq\E\dab\Bigr).
	\end{align}
\end{subequations}
The lattice operators (\ref{eq:opCombi_fi}, \ref{eq:moment_functionals_g_Rabol}) are the final results for the kinetic scheme. They define the ``perturbation'' process and can be implemented after the semi-Lagrangian advection step to successively simulate the spatio-temporal evolution of $\rho, \bmu, E$ and other macroscopic quantities of interest. 

\subsection{Thermodynamics}
\label{sec:Thermodynamics}
Phase transformations include non-equilibrium thermodynamics and large variations in viscosity and density that reside on a multitude of spatio-temporal scales that historically have been difficult to treat in the continuum without resorting to limiting assumptions that appear in many empirical mass-transfer models for multi-component mixtures, as we have documented in a recent review \cite{Petersen2023a}. 

Whereas mixture models account for the latent heat transfer between species $\dot{m}h_\textrm{lv}$, our current single-component implementation evolves $T,p$ and phase transitions under the saturation curve in tandem with the empirical van der Waals (VDW) equation of state (EOS). Even though the EOS is empirical its simple non-monotonic topology with a distinct energy minimum enables nucleation in kinetic theories as was shown in \cite{Reyhanian2020} that employ the same EOS. Moreover, the cubic VDW-EOS is the simplest EOS that retains the ability to predict spinodals \cite{Aursand2017} and is inherently mechanically unstable in the spinodal region and metastable in the binodal region. When a metastable fluid is perturbed the fluid state will transition towards the binodal curve where the local free energy is minimized \cite{Hosseini2023}. \citet{Reyhanian2020} presented the governing thermodynamic equations, but for conciseness we reproduce them here for later reference in our multi-scale analysis (\S\ref{sec:CEexpansion}). We duly note that the kinetic equations conserve the density \eqref{eq:rho}, momentum \eqref{eq:ja}, and total energy \eqref{eq:E}, where the latter predicts the internal energy $e$ via,
\begin{align}
	E	= e + \frac{\ua\ua}{2}.
\end{align}
Thus, given $e$, the temperature can be computed by reorganizing the EOS, 
\begin{align}
	\label{eq:evdw}
	e 	= c_v T - a\rho - \frac{p_0}{\rho},
\end{align}
in which the term $p_0/\rho$ is included to retain positive temperatures in the subcritical region $T/T_\textrm{cr} < 0.84375$ \cite{Reyhanian2021a}. We can then proceed with defining the thermodynamic pressure as, 
\begin{align}
	\label{eq:pvdw}
	p 	= \frac{\rho RT}{1 - b\rho} - a\rho^2,
\end{align}
which contributes the non-ideal energy contribution $T\bigl(\partial p/\partial T\bigr)_v \parta\ua$ in the Fourier equation, as we will document later in the Chapman-Enskog multiscale analysis in \S\ref{sec:Moment equations}. This analysis further necessitates deriving the isobaric specific heat and speed-of-sound, to arrive at the thermo-hydrodynamic limit. The former reads,
\begin{align}
	\specificcp \doteq \derhTp,
\end{align}
and the real-gas speed-of-sound, 
\begin{align}
	\cssq \doteq \derprhos,
\end{align}
and can be rewritten using the cyclic and Maxwell relations \cite{Borgnakke2013} as, 
\begin{align}
	\specificcp 	&= \dereTp + p\dervTp, \\
	\label{eq:cssq}
	\cssq 			&= \derprhoT + \frac{T}{\rho^2\cv}\derpTv^2.
\end{align}

\section{Chapman-Enskog multiscale analysis}
\label{sec:CEexpansion}
As it is our goal to utilize the kinetic model \eqref{eq:LFPBEs} in mesoscale simulations where we recover macroscopic quantities (\ref{eq:continuous_moments_f}, \ref{eq:continuous_moments_g}) from microscopic populations, it is crucial to investigate how the equations behave across scales. Specifically, we conduct a Chapman-Enskog multiscale analysis \cite{Kruger2017}. Initially, the post-advection state is approximated by a Taylor series, after which the modified kinetic equations are expanded in terms of perturbation series of all the variables. Subsequently, all terms in the equations can be separated by their representative orders forming a hierarchy of perturbation relations for the advection and collision sides of the equations. Then, the zero-through-second moments can be taken of these relations to form moment equations, i.e. scale-dependent forms of the continuum conservation laws. This step is crucial as it comprises the moments stemming from non-equilibrium, higher-order contributions of the perturbation series that can be chosen to either be included or excluded with the aim of assessing their mathematical formalism and physical implications in the continuum, thus creating a direct avenue between the microscopic and macroscopic scales. Finally, the Navier-Stokes-Fourier (NSF) equations can be obtained by recombining the perturbation series. By including selected non-equilibrium populations in the moment equations it is possible to elucidate the asymptotic behavior of different parts of the NSF hierarchy, to which end we initially focus on the shear-stress tensor.    

\subsection{Perturbation series}
The mesoscale behavior, especially that stemming from the perturbations in the FPE, evolves at different characteristics length and time scales. Analysis of the scale-effects as they manifest in the continuum can be treated with multiscale methods in which we seek to separate derivative terms, populations, and macroscopic observables into contributions that scale with a ``smallness'' parameter $\e$ where $\e \ll 1$ is associated with the thermo-hydrodynamic limit \cite{Succi2018}. By exemplification, the viscous effects of the FPE may reside on longer time scales whereas shocks propagate on shorter time scales. Specifically, we decompose our distributions $f_i,g_i$, spatial and temporal gradients $\parta,\partt$, and macroscopic moments into a family of perturbation series around $\e$ raised to different exponent values. Firstly, the populations expand to, 
\begin{subequations}
	\label{eq:perturbation_series}
	\begin{align}
		f_i 	&= \fizero + \e\fione + \e^2\fitwo + \mathO\bigl(\e^3\bigr),\\
		g_i 	&= \gizero + \e\gione + \e^2\gitwo + \mathO\bigl(\e^3\bigr),
	\end{align}
	where the zeroth-order contributions are the known equilibrium distributions, such that the higher orders act as unknown, increasingly non-linear components. Propagation and diffusion phenomena are captured by decomposing the time derivative into two time variables,
	\begin{align}
		\partt 	= \e\parttone + \e^2\partttwo + \mathO\bigl(\e^3\bigr). 
	\end{align}
	As for the spatial derivatives, these are evolved on the hydrodynamic scale that is appropriately resolved as, 
	\begin{align}
		\parta	= \e\partaone + \mathO\bigl(\e^2\bigr). 
	\end{align}
	Moreover, we are interested in the zeroth-through-second-order perturbations in all of the macroscopic observables in order to recover the correct continuum-conservation laws, 
	\begin{gather}
		\label{eq:perturbation_series_rho}
		\rho 	= \rho\superzero + \e\rho\superone + \e^2\rho\supertwo,\\ 
		\E 		= \E\superzero + \e\E\superone + \e^2\E\supertwo,\\
		\ja 	= \ja\superzero + \e\ja\superone + \e^2\ja\supertwo,\\ 
		\qa 	= \qa\superzero + \e\qa\superone + \e^2\qa\supertwo,\\
		\Pab 	= \Pab\superzero + \e\Pab\superone + \e^2\Pab\supertwo,\\
		\label{eq:perturbation_series_Rab}
		\Rab 	= \Rab\superzero + \e\Rab\superone + \e^2\Rab\supertwo,
	\end{gather}
	and similarly for the linear recombinations of the above, $\rhool = \rhool\superzero + \e\rhool\superone + \e^2\rhool\supertwo$ and so forth. 
\end{subequations}
We interpret the zeroth orders as ``equilibrium'' contributions that dictate the phenomena that evolve on the macroscales, whereas increasing orders represent perturbations on faster time and smaller length scales, respectively. Consequently, we can imagine that microscopic fluctuations due to thermal noise may be observable on the $\e^1,\e^2$ scales, where the coarser fluctuations associated with $\e^1$ are likely more relevant to analyze in the context of mesoscopic nucleation.    

\subsection{Expansions: Taylor series, perturbations, separation, moments, and recombination}
\subsubsection{Perturbation-series expansion}
We expand the previously derived lattice equations \eqref{eq:LFPBEs} to prove their consistency in the thermo-hydrodynamic limit. In the following we adopt the notation for the pre-advection populations $f_i = f_i(\bmx - \bmvi\dt, t-\dt)$ and post-advection populations $\fiprm = f_i(\bmx, t)$. First we treat the advection side of the equations, and seek to approximate the advected state $\fiprm$ by its Taylor expansion around the datum $f_i$ where we exploit the common \textit{ansatz} that third and higher order terms $n > 2$ are very small and do not significantly affect the macroscopic behavior \cite{Kruger2017}. As such, to find the Navier-Stokes-Fourier equations, we only retain the two lowest orders in the \textit{Taylor-series} expansion, 
\begin{align}
	\fiprm 	= f_i &+ \dt \bigl(\partt + \vid\partd\bigr)f_i \nonumber\\
	&+ \frac{\dt^2}{2}\bigl(\partt + \vid\partd\bigr)\bigl(\partt + \vie\parte\bigr) f_i + \mathcal{O}\bigl(\dt^3\bigr),
\end{align}
and similarly for $g_i$. It follows that the lattice equations \eqref{eq:LFPBEs} can be rewritten to, 
\begin{align}
	\dt \bigl(\partt &+ \vid\partd\bigr)f_i + \frac{\dt^2}{2}\bigl(\partt + \vid\partd\bigr)\bigl(\partt + \vie\parte\bigr) f_i \nonumber\\
	&= \omgFP\opFPi + \omgBGK\opBGKi,
\end{align}
where we have neglected the higher-order $\mathcal{O}\bigl(\dt^3\bigr)$ terms. Before treating the collision side of the equation, we make the initial perturbation analysis on the advection terms and retain them for further analysis. The analysis starts with another common \textit{ansatz}: only the two lowest orders \add{in the Knudsen number $Kn \equiv \overline{\lambda}/L_c$ (where $\overline{\lambda}$ is the molecular mean-free path and $L_c$ the pertinent characteristic length scale)} are required to obtain the continuum NSEs where only the coarse non-equilibrium effects in $\fione$ are considered in addition to the equilibrium $\fizero$ \cite{Kruger2017}. Substituting in the pertinent variables from \eqref{eq:perturbation_series} yields the expanded advection side,
\begin{widetext}
	\begin{align}
		\biggl\{\dt \Bigl[\bigl(\e\parttone + \e^2\partttwo\bigr) + \e\via\partaone\Bigr] + \frac{\dt^2}{2}\Bigl[\bigl(\e\parttone + \e^2\partttwo\bigr) + \e\via\partaone\Bigr]\Bigl[\bigl(\e\parttone + \e^2\partttwo\bigr) + \e\vib\partbone\Bigr]\biggr\}\bigl(\fizero + \e\fione + \e^2\fitwo\bigr),
	\end{align} 
\end{widetext}


\noindent where contributions \textit{separated} by orders in $\e$ are,
\begin{subequations}
	\label{eq:eorders_advection}
	\begin{align}
		&\e^0:\qquad 	0,\\
		&\e^1:\qquad 	\dt \bigl(\parttone + \via\partaone\bigr)\fizero,\\
		&\e^2:\qquad 	\Bigl[\dt\partttwo + \frac{\dt^2}{2}\bigl(\parttone + \via\partaone\bigr)\bigl(\parttone + \vib\partbone\bigr)\Bigr]\fizero\nonumber\\ 
		&\qquad\qquad + \dt\Bigl(\parttone + \via\partaone\Bigr)\fione.
	\end{align}
\end{subequations}
These contributions will eventually equate with the corresponding-order contributions from the analogously expanded collisions operators. To that end, the BGK operator is expanded with the perturbation series yielding the separated orders, 
\begin{subequations}
	\label{eq_eorders_BGK}
	\begin{align}
		&\e^0:\qquad 	\omgBGK\bigl(\fieq - \fizero\bigr),\\
		&\e^1:\qquad 	-\omgBGK\fione,\\
		&\e^2:\qquad 	-\omgBGK\fitwo.
	\end{align}
\end{subequations}
Proceeding with the FP operator, we similarly expand it by parts in $\opFPi = \opFPizero + \e\opFPione + \e^2\opFPitwo$ (wherein the perturbation order $n_{\e}$ should not be confused with those of the Hermite series), in which each of the orders are further decomposed by $n_{\e} = \{0,1,2\}$, 
\begin{align}
	\opFPine = w_i\left[\rhoolen + \frac{\via}{\vTsq}\Jaolen + \frac{\via\vib - \vTsq\dab}{2\vT^4}\Pabolen\right].
\end{align}
Thus, separation of orders simply yields, 
\begin{subequations}
	\label{eq_eorders_FP}
	\begin{align}
		&\e^0:\qquad 	\omgFP\opFPizero,\\
		&\e^1:\qquad 	\omgFP\opFPione,\\
		&\e^2:\qquad 	\omgFP\opFPitwo,
	\end{align}
\end{subequations}
where the various functionals \eqref{eq:moment_functionals_f_rhool}--\eqref{eq:moment_functionals_g_Rabol} are linear combinations of the perturbation series of the macroscopic variables \eqref{eq:perturbation_series_rho}--\eqref{eq:perturbation_series_Rab}, such that the corresponding $\e$-orders are also linearly retained. 

Now, the next step is to recombine the different orders of the perturbation series of the advection terms, and the BGK and FP operators, resulting in, 
\begin{subequations}
	\begin{align}
		\label{eq:zerothorder_perturbation_equation}
		&\e^0:\qquad 	0 = \omgBGK\bigl(\fieq - \fizero\bigr) \nonumber\\
		&\qquad +  \omgFP w_i\left[\rhoolzero + \frac{\via}{\vTsq}\jaolzero + \frac{\via\vib - \vTsq\dab}{2\vT^4}\Pabolzero\right],\\
		\label{eq:firstorder_perturbation_equation}
		&\e^1:\qquad	\bigl(\parttone + \vid\partdone\bigr)\fizero = -\frac{\omgBGK}{\dt}\fione\nonumber\\  
		&\qquad + \frac{\omgFP}{\dt} w_i\left[\rhoolone + \frac{\via}{\vTsq}\Jaolone + \frac{\via\vib - \vTsq\dab}{2\vT^4}\Pabolone\right],\\
		&\e^2:\qquad 	\Bigl[\partttwo + \frac{\dt}{2}\bigl(\parttone + \vid\partdone\bigr)\bigl(\parttone + \vie\parteone\bigr)\Bigr]\fizero \nonumber\\ 
		&\qquad + \Bigl(\parttone + \vid\partdone\Bigr)\fione =  -\frac{\omgBGK}{\dt}\fitwo\nonumber\\
		\label{eq:secondorder_perturbation_equation_v0}
		& + \frac{\omgFP}{\dt} w_i\left[\rhooltwo + \frac{\via}{\vTsq}\Jaoltwo + \frac{\via\vib - \vTsq\dab}{2\vT^4}\Paboltwo\right],
	\end{align}
\end{subequations}
to which end we make a third \textit{ansatz}: there exists multiple roots in the zeroth-order equality \eqref{eq:zerothorder_perturbation_equation} where $\fizero = \fieq$ is a candidate as is conventionally the case in multi-scale analyses of the Boltzmann equation. We can further infer that for the equality to hold, we must require that, 
\begin{align}
	\rhoolzero 	&= 0, \\
	\jaolzero 	&= 0\suba, \\
	\Pabolzero 	&= 0\subab.	
\end{align}
The zero equalities are the only information we can extract from the zeroth-order relation for now. Prior to taking the moments of the perturbation relations, we can simplify the first and second-order relations. If we factorize $\fione$ out of $\opFPione$ in \eqref{eq:firstorder_perturbation_equation} and adopt the notation $\opHatFPione \circ \fione$ as the FP operational expression operating on the first-order populations, the material derivative of the equilibrium populations can be written as, 
\begin{align}
	\label{eq:firstorder_perturbation_equation_v1}
	\e^1:\qquad 	\tottbmvone\fizero 	= \fione \left(\frac{\omgFP}{\dt}\opHatFPi - \frac{\omgBGK}{\dt}\right),
\end{align}
where we employ the notation $\tottbmvone = \parttone + \vid\partdone$ for conciseness, and recognize that $\opHatFPi$ is in fact independent of the perturbation order. This result is useful for eliminating the first-order derivatives of $\fizero$ in \eqref{eq:secondorder_perturbation_equation_v0}, and combining all prefactors of $\fione$, by substituting in this reformulated $\e^1$ contribution yielding,
\begin{align}
	\label{eq:secondorder_perturbation_equation_v1}
	\e^2:\qquad 	&\partttwo\fizero + \tottbmvone\left(1 + \frac{\omgFP\opHatFPi}{2} - \frac{\omgBGK}{2}\right)\fione \nonumber\\ 
	&= -\frac{\omgBGK}{\dt}\fitwo + \frac{\omgFP}{\dt} \opFPitwo.
\end{align} 
This allows us to again in a reverse manner recast the relation with $\fione$ isolated in \eqref{eq:firstorder_perturbation_equation_v1}, such that the second-order contributions now obey, 
\begin{align}
	\label{eq:secondorder_perturbation_equation_v2}
	\e^2:\qquad 	&\partttwo\fizero + \Bigl(\parttone + \vid\partdone\Bigr)\Bigl(\parttone + \vie\parteone\Bigr)\dt\nonumber\\
	&\times\left(1 + \frac{2 - \omgBGK}{2\omgFP\opHatFPi} - \frac{2 + \omgFP\opHatFPi}{2\omgBGK}\right)\fizero \nonumber\\
	&\qquad = -\frac{\omgBGK}{\dt}\fitwo + \frac{\omgFP}{\dt} \opFPitwo,
\end{align} 
where we have obtained an equilibrium differential equation with $\fizero$ instead of $\fione$, where the non-equilibrium effects are administered through the $\omgBGK,\omgFP$ coefficients of $\fizero$. The perturbation equations \eqref{eq:zerothorder_perturbation_equation}, \eqref{eq:firstorder_perturbation_equation_v1}, \eqref{eq:secondorder_perturbation_equation_v2} are now at a stage where we can further the analysis by forming the moment equations.

\subsubsection{Moment equations}
\label{sec:Moment equations}
Initially we can take the zero, second, and third moments of \eqref{eq:firstorder_perturbation_equation}, from which the continuity and momentum equation are obtained. The zeroth-moment equation is obtained by summing both sides of the relation over $i$, the first-moment equation by summing over $i$ multiplied by $\vie$, the second-moment equation by $\vie\viz$, and so forth. Thereto, we can investigate the effects of perturbations on the different macroscopic observables by including or excluding various moments. In the simplest analysis, we can make the \textit{ansatz} that the non-equilibrium moments \cite{Chapman1952},
\begin{subequations}
	\label{eq:noneq_moment_identities}
	\begin{gather}
		\label{eq:noneq_zeromoment_identity}
		\sumQ \bigl\{\fineq, \gineq\bigr\} = \bigl\{\rhoneq, \Eneq\bigr\} = 0, \\
		\label{eq:noneq_firstmoment_identity}
		\sumQ \bigl\{\fineq, \gineq\bigr\} \vid = \bigl\{\jdneq, \qdneq\bigr\} = 0\subd,\\
		\label{eq:noneq_secondmoment_identity}
		\sumQ \bigl\{\fineq, \gineq\bigr\} \vid\vie = \bigl\{\Pdeneq, \Rdeneq\bigr\} = 0\subde.
	\end{gather}
\end{subequations} 
are all zero for the non-equilibrium states in $n_{\e} \ge 1$. This conventionally renders the continuity and Euler equations, and does not give any insight into non-equilibrium effects. Now, as we are interested in investigating the non-equilibrium behavior of the FP equation on smaller scales it may be worthwhile to contest the conventional zero-assumptions of (\ref{eq:noneq_zeromoment_identity}--\ref{eq:noneq_secondmoment_identity}) and include the coarse first-order perturbations in the analysis. As we know that the entropy-extremum principle corresponds to a minimization of energy \cite{Callen1985}, we can include the contributions from $\Eone = \sumi\gione$. Additionally, for thermally-driven phase-change processes such as pool boiling it is apt to consider the fluctuations in the heat flux $\qdone = \sumi\gione\vid$, and for cavitation which is a stress-induced phenomenon, the momentum flux $\Pdeone = \sumi\fione\vid\vie$. As the inclusion of $\Pdeone$ in the second-moment equation of the $\e$-relation based on the Boltzmann equation results in the viscous-stress tensor $\taude$ with the dynamic viscosity $\mu = \bigl(1/\omgBGK - 1/2\bigr)p\dt$ \cite{Reyhanian2020}, one could suspect that $\omgFP$ \eqref{eq:omgFP} could be attributed to a pseudo-viscosity from the $\gamma$-damping of thermal fluctuations---analogous to the eddy-viscosity in turbulence models. Thus, we are interested in the same non-equilibrium moments and how they are affected by the FP dynamics, and as such \textit{assume} that the values and correlations of all non-equilibrium moments, except for that \textit{unknown} and \textit{non-zero} $\Pdeone$ moment, are zero, 
\begin{subequations}
	\label{eq:noneq_moments_specified}
	\begin{align}
		\rhooltwo 													&\doteq 	0,\\
		\jaone 		= \jatwo 		= \Jaolone 		= \Jaoltwo		&\doteq 	0\suba,\\
		\Pabtwo  	= \Paboltwo										&\doteq 	0\subab,\\
		\Eone 		= \Etwo 		= \Eolone 		= \Eoltwo 		&\doteq 	0,\\
		\qatwo 		= \qaoltwo 										&\doteq 	0\suba,\\
		\Rabone 	= \Rabtwo  		= \Rabolone 	= \Raboltwo 	&\doteq 	0\subab,
	\end{align}
\end{subequations}
in addition to the non-zero counterpart, 
\begin{subequations}
	\label{eq:nonzero_noneq_moments_specified}
	\begin{align}
		\Pabone 	\neq 0.
	\end{align}
\end{subequations}
This ultimately allows us to take the zero-through-second moments of (\ref{eq:firstorder_perturbation_equation}, \ref{eq:secondorder_perturbation_equation_v1}), which usually are necessary to recover the NS equations. Moreover, for the FP contributions, we are primarily interested in the non-equilibrium effects to the momentum and energy equations, and as such we omit the density contributions in the zero-moment continuity equation. As for the Boltzmann collision operator, we rely on the usual methodology of omitting non-equilibrium contributions from the zero and first moment equations of the perturbation equations \cite{Kruger2017}. From the moments of the first-order perturbation expansion we get the (first-order) continuity, Euler, and a higher-order equation, 
\begin{subequations}
	\begin{align}
		\label{eq:zero_moment_eq}
		\parttone\rho\supereq + \partdone\jdeq &= 0, \\
		\parttone\jdeq + \parteone\Pdeeq &= 0,\\
		\label{eq:second_moment_eq}
		\parttone\Pdeeq + \partzone\Qdezeq &= -\frac{\omgBGK}{\dt}\Pdeone + \frac{\omgFP}{\dt}\opHatFP\Pdeone,
	\end{align}
	in which we exploited the zero-non-equilibrium moments of $\fione$, as well as factorized out $\sumQ f_i$ from $\opFP$ leaving the moment of the operational expression itself $\opHatFP$. We will show later that the last moment equation, as in the LBM, recovers the viscous-stress tensor, albeit with additional stresses manifested by $\opHatFP\Pdeone$. To that end, the explicit formalism of $\Pdeone$ is not recovered directly from \eqref{eq:second_moment_eq} but rather appears in the first-moment of the second-order perturbation expansion \eqref{eq:secondorder_perturbation_equation_v2}. Lastly, the equilibrium moments are defined from the equilibrium populations \eqref{eq:fieq} as, 
	\begin{gather}
		\rho\supereq 	= \sumQ f_i\supereq 				= \rho,\\
		\jdeq 			= \sumQ f_i\supereq \vid 			= \rho\ud,\\
		\Pdeeq 			= \sumQ f_i\supereq \vid\vie 		= \rho\ud\ue + p\dde,\\
		\Qdezeq 		= \sumQ f_i\supereq \vid\vie\viz 	= \rho\ud\ue\uz + p\bigl[u\delta\bigr]\subdez,
	\end{gather}
\end{subequations}
where $\bigl[u\delta\bigr]\subdez = \ud\dez + \ue\ddz + \uz\dde$. Applying the product rule to $\partdone(\rho\ud)$ in the continuity equation \eqref{eq:zero_moment_eq} enables rewriting the material derivative as, 
\begin{align}
	\label{eq:Dt_rho}
	\Dtuone\rho 	= - \rho\partdone\ud. 
\end{align}
This material derivative appears in the product-rule expanded momentum equation considering the equilibrium moments $\jdeq$ and $\Pdeeq$, which after some algebra allows us to rewrite the momentum equation as, 
\begin{align}
	\label{eq:Dt_u}
	\Dtuone\ud 	= -\frac{1}{\rho}\parteone p \dde.
\end{align}
Noting that the same perturbation series apply to $g_i$, its corresponding moment equation hierarchy is, 
\begin{subequations}
	\begin{gather}
		\parttone\E\supereq + \partdone\qdeq = 0,\\
		\parttone\qdeq + \parteone\Rdeeq = -\frac{\omgBGK}{\dt}\qdone + \frac{\omgFP}{\dt}\opHatFP\qdone, 
	\end{gather}
	where the moments of the equilibrium populations \eqref{eq:gieq} read,
	\begin{gather}
		\E\supereq 		= \sumQ g_i\supereq 			= 2\rho E,\\
		\qdeq 			= \sumQ g_i\supereq\vid 		= 2\rho\ud H, \\
		\Rdeeq			= \sumQ g_i\supereq\vid\vie	= 2\rho\ud\ue\bigl(H + p/\rho\bigr) + 2pH\dde,
	\end{gather}
	and $H = E + p/\rho = e + u^2/2 + p/\rho$ is the total enthalpy. All of the equilibrium moments except $\Qdezeq$ are symmetric tensors. To arrive at the NSF equations, the second $g_i$-moment equation is not required, and thus we have excluded it. 
\end{subequations}
After some expansion with the product rule, and substituting the van der Waals internal energy \eqref{eq:evdw}, the derivative relation $T\bigl(\partial p/\partial T\bigr)_v = a\rho^2 + p$ on a VDW-EOS basis, and the rewritten momentum equation \eqref{eq:Dt_rho}, it can be shown that the first-order contribution to the energy equation is,
\begin{align}
	\label{eq:Dt_T}
	\Dtuone T 	= - \frac{T}{\rho C_v}\derpTv\partdone\ud. 
\end{align}
Furthermore, an analogous pressure equation can be established by considering that $p = p(\rho,T)$ where applying the corollary chain rule,
\begin{align}
	\Dtuone p 	= \derprT\Dtuone\rho + \derpTr\Dtuone T,
\end{align}
can be rewritten with the first-order continuity and energy equation contributions (\ref{eq:Dt_rho},\ref{eq:Dt_T}), as well as the speed-of-sound \eqref{eq:cssq}, yielding,  
\begin{align}
	\label{eq:Dt_p}
	\Dtuone p 	= -\rho\cssq\partdone\ud.
\end{align}
This relation will be exploited later to derive the viscous stresses in the moment equations. As of now we have obtained the necessary zero and first-moment equations of $f_i,g_i$, and are now left with deriving an operational expression for the moment $\Pdeone$ terms in the second-moment equation \eqref{eq:second_moment_eq}. This last part of the analysis revolves around the second-order perturbations \eqref{eq:secondorder_perturbation_equation_v1} from which the zero moment becomes,  
\begin{align}
	\label{eq:dttwo_rho}
	\partttwo\rho 	&= 0,
\end{align}
and consequently we can conclude that the density predominantly evolves on the $\epsilon$ scales via \eqref{eq:Dt_rho}. Formulating the first-moment equation of the $\epsilon^2$ scales necessitates more algebraic work. First we note that in the expanded first-moment equation stemming from \eqref{eq:secondorder_perturbation_equation_v2} successive instances of the momentum equation \eqref{eq:Dt_u} appear and significantly simplify the result to, 
\begin{align}
	\partttwo \rho \ud 		= \parteone \left(\parttone\Pdeeq + \partzone\Qdezeq\right)\dt\biggl[\left(\frac{1}{\omgBGK} - 1\right)\nonumber\\ 
	+ \left(\frac{\omgFP}{2\omgBGK}\opHatFP - \frac{1 - \omgBGK/2}{\omgFP}\frac{1}{\opHatFP}\right)\biggr],
\end{align}
where the moment $\opHatFP$ of the operator is $\sumQ \opHatFPi$. Nevertheless, given that $\opHatFP = \opHatFP(\bmx,\bmu, t)$ we cannot immediately exclude the relaxation-frequency-dependent terms, except for $\bigl(1/\omgBGK - 1\bigr)$, in the derivative terms, and thus need to dissect the equation further. After some algebra it follows that the second-order contributions to the momentum equation take the form, 
\begin{widetext}
	\begin{align}
		&\partttwo\ud 	= \frac{\dt}{\rho}\biggl\{\left(\frac{1}{\omgBGK} - 1\right)\parteone\left(\parttone\Pdeeq + \partzone\Qdezeq\right)\nonumber\\
		&\quad + \frac{\omgFP}{2\omgBGK}\parteone\Bigl[\opHatFP\left(\parttone\Pdeeq + \partzone\Qdezeq\right) + \left(\Pdeeq\parttone + \Qdezeq\partzone\right)\opHatFP\Bigr]\nonumber\\
		&\quad - \frac{1 - \omgBGK/2}{\omgFP}\parteone\biggl[\left(\opHatFP\right)^{-1}\left(\parttone\Pdeeq + \partzone\Qdezeq\right) + \left(\Pdeeq\parttone + \Qdezeq\partzone\right)\left(\opHatFP\right)^{-1}\biggr]\biggr\},
		\label{eq:dttwo_u}
	\end{align} 
\end{widetext}
in which the superposition of the three prefactors of $\parteone(\ \cdots)$ recover the dynamic viscosity in the thermo-hydrodynamic limit that together with $\parttone\Pdeeq + \partzone\Qdezeq$ constitute the viscous-stress tensor, as we will show in the following. In the result we can observe three major components to the stresses: the ordinary Boltzmann-equation contributions by $\bigl(1/\omgBGK - 1\bigr)$ as well as contributions from the FP operator with both the material derivative of itself (either in the continuous \eqref{eq:FPE_cont} or derived Hermite-space formalism) linearly dependent on the moments, as well as more FPE-analogous material derivatives of the moments operated on by the operator.

In an analogous analysis of the perturbation expansion of $g_i$, only the zero-moment is required for recovering the Fourier equation. Considering the perturbation expansion of not $f_i$ but $g_i$ in \eqref{eq:secondorder_perturbation_equation_v2} the resulting moment equation simplifies to, 
\begin{align}
	&\partttwo T 	= \frac{1}{2\rho C_v}\biggl\{\partdone\left(\parttone\qdeq + \parteone\Rdeeq\right)\dt\biggl[\left(\frac{1}{\omgBGK} - 1\right)\nonumber\\ 
	&\quad + \left(\frac{\omgFP}{2\omgBGK}\opHatFP - \frac{1 - \omgBGK/2}{\omgFP}\frac{1}{\opHatFP}\right)\biggr] - 2\rho\ud\partttwo\ud\biggr\},
\end{align} 
which requires exploiting the thermodynamic pressure \eqref{eq:pvdw}, the internal energy \eqref{eq:evdw}, the continuity equation \eqref{eq:Dt_rho}, and the aforementioned pressure equation \eqref{eq:Dt_p}. Finally, $\opHatFP = \opHatFP(\bmx,\bmu, t)$, necessitates further expansion yielding,
\begin{widetext}
	\begin{align}
		&\partttwo T 	= \frac{1}{2\rho C_v}\biggl\{\dt\left(\frac{1}{\omgBGK} - 1\right)\partdone\left(\parttone\qdeq + \parteone\Rdeeq\right) - 2\rho\ud\partttwo\ud \nonumber\\
		&\quad + \dt\left(\frac{\omgFP}{2\omgBGK}\right)\partdone\left[\opHatFP\left(\parttone\qdeq + \parteone\Rdeeq\right) + \left(\qdeq\parttone + \Rdeeq\parteone\right)\opHatFP\right] \nonumber\\
		&\quad - \dt\left(\frac{1 - \omgBGK/2}{\omgFP}\right)\partdone\left[\left(\opHatFP\right)^{-1}\left(\parttone\qdeq + \parteone\Rdeeq\right) + \left(\qdeq\parttone + \Rdeeq\parteone\right)\left(\opHatFP\right)^{-1}\right]\biggr\}.
		\label{eq:dttwo_T}
	\end{align}
\end{widetext}
This hierarchy of equations---specifically the continuity (\ref{eq:Dt_rho},\ref{eq:dttwo_rho}), the momentum (\ref{eq:Dt_u},\ref{eq:dttwo_u}), and the energy (\ref{eq:Dt_T},\ref{eq:dttwo_T}) equations---form the foundation for recovering the NSF equations by \textit{recombination} of the perturbation components into the native series formulation \eqref{eq:perturbation_series}. 

\subsubsection{Thermo-hydrodynamic limit:\\ the Navier-Stokes-Fourier equations}
At this point, we recover the Navier-Stokes-Fourier equations from the hierarchy of perturbed moment equations by recombining each of the orders in $\e$. By merging (\ref{eq:Dt_rho},\ref{eq:dttwo_rho}), and recognizing that $\e\parttone + \e^2\partttwo = \partt$ and $\e\partdone = \partd$ we get the continuity equation, 
\begin{align}
	\partt\rho + \ud\partd\rho = -\rho\partd\ud.
\end{align}
Similarly, the momentum equation becomes, 
\begin{subequations}
	\begin{align}
		\rho\Dtu\ud 	&= -\parte p\dde - \parte\bigl(\taude + \taudetld\bigr),
	\end{align}
	where the viscous-stress tensor stemming exclusively from Boltzmann-type collisions is, 
	\begin{align}
		\label{eq:taude}
		\taude 	= -\muB\left(\partd\ue + \parte\ud - \frac{2}{D}\bigl(\partz\uz\bigr)\dde\right)\nonumber\\ 
		- \zB\bigl(\partz\uz\bigr)\dde.
	\end{align}
	In addition to this ``traditional'' stress tensor, we also get ``noisy'' stress-tensor components perturbed directly by the FP operator and its inverse, separated by complementing \textit{perturbed} components (denoted by ``$+$'' and ``$-$'' subscripts) that concurrently amplify and attenuate stresses \add{(assuming $\opHatFP > 1$)}, 
	\begin{widetext}
		\begin{align}
			\label{eq:taudetld}
			\taudetld 	&= \overbrace{\opHatFP\left[-\muFPBpls\left(\partd\ue + \parte\ud - \frac{2}{D}\bigl(\partz\uz\bigr)\dde\right) - \zFPBpls\bigl(\partz\uz\bigr)\dde\right]}^{(\textrm{I})}\nonumber\\
			&+ \frac{1}{\opHatFP}\left[-\muFPBmns\left(\partd\ue + \parte\ud - \frac{2}{D}\bigl(\partz\uz\bigr)\dde\right) - \zFPBmns\bigl(\partz\uz\bigr)\dde\right]\hspace{5.6cm} \Biggr\}{\footnotesize(\textrm{II})}\nonumber\\
			&+ \muFPBpls\frac{\Pdeeq}{p}\left\{\left[\left(\partial_{\vzeta}\opHatFP\right)\frac{A\subz\rho\cssq}{\gamma m} + \left(\partial_{\vTsq}\opHatFP\right)\frac{k_B T}{m\rho\cv}\derpTv \right]\bigl(\partz\uz\bigr) + \left(\partial_{\uz}\opHatFP\right)\frac{1}{\rho}\bigl(\partd p \bigr)\ddz \right\}\qquad\Biggr\}{\footnotesize(\textrm{III})}\nonumber\\
			&+ \underbrace{\muFPBmns\frac{\Pdeeq}{p}\left\{\left[\left(\partial_{\vzeta}\frac{1}{\opHatFP}\right)\frac{A\subz\rho\cssq}{\gamma m} + \left(\partial_{\vTsq}\frac{1}{\opHatFP}\right)\frac{k_B T}{m\rho\cv}\derpTv \right]\bigl(\partz\uz\bigr) + \left(\partial_{\uz}\frac{1}{\opHatFP}\right)\frac{1}{\rho}\bigl(\partd p \bigr)\ddz \right\}}_{(\textrm{IV})}.
		\end{align}
	\end{widetext}
	In (\ref{eq:taude},\ref{eq:taudetld}) the viscosities manifest due to both the Boltzmann and FP relaxation frequencies,
	\begin{gather}
		\muB 	= \left(\frac{1}{\omgBGK} - 1\right)p\dt,\\
		\zB 	= \left(\frac{1}{\omgBGK} - 1\right)\left(\frac{D + 2}{D} - \frac{\rho\cssq}{p}\right)p\dt,
	\end{gather}
	\begin{gather}
		\muFPBpls 	= \left(\frac{\omgFP}{2\omgBGK}\right)p\dt,	\nonumber\\		
		\muFPBmns 	= \left(-\frac{1 - \omgBGK/2}{\omgFP}\right)p\dt,\nonumber\\
		\zFPBpls 	= \left(\frac{\omgFP}{2\omgBGK}\right)\left(\frac{D + 2}{D} - \frac{\rho\cssq}{p}\right)p\dt, \nonumber\\		
		\zFPBmns = \left(-\frac{1 - \omgBGK/2}{\omgFP}\right)\left(\frac{D + 2}{D} - \frac{\rho\cssq}{p}\right)p\dt. 
	\end{gather}	
\end{subequations}
Thus, we get four pair-wise complementing, \add{coarse-grained ``stochastic''} stress tensors in \eqref{eq:taudetld}: two traditional stress tensor components (I,II) pre-factored by FP-operator $\opHatFP$ and its inverse counterpart $1/\opHatFP$. The remaining two viscous-stress components (III,IV) are due to the spatial and temporal derivatives of $\opHatFP$ in \eqref{eq:taudetld}, and account for the diffusion contributed from the FP operator itself as it varies due to external forces (via $\bmveta$ in the $\partial_{\vzeta}$ terms), thermal fluctuations (via $\vTsq$ in the $\partial_{\vTsq}$ terms), and continuum velocity ($\bmu$ in the $\partial_{\uz}$ terms), respectively. We duly note that by \add{\textit{stochastic}} we imply that the dynamics of the Langevin equation is partially captured in the coarse-grained FPE rather than computed directly as discrete, stochastic events sampled at random from a characteristic Gaussian white noise distribution, and that the coarse-grained noise permeates the stresses. \add{We use the term stochastic rather than ``fluctuating'' to not confuse our methodology with the aforementioned fluctuating-hydrodynamics theory originating from \citet{Landau1956a}}.

In obtaining the velocity gradient and divergence terms in the stresses, i.e. $\left(\partd\ue + \parte\ud - 2/D\bigl(\partz\uz\bigr)\dde\right)$ and $\bigl(\partz\uz\bigr)\dde$, we have exploited the expanded formulation, 
\begin{align}
	\parttone\Pdeeq &+ \partzone\Qdezeq 	= p \left(\partdone\ue + \parteone\ud\right) \nonumber\\
	&+ \left(p - \rho\cssq\right)\partzone\uz\dde,
\end{align}
previously identified in \eqref{eq:dttwo_u}. The noisy stresses arise analogously to the stresses by the Boltzmann operator. During the recombination we obtain the auxiliary components with the derivatives $\left(\Pdeeq\parttone + \Qdezeq\partzone\right)\Bigl\{\opHatFP;1/\opHatFP\Bigr\}$ imposed on the FP operational expressions and its inverse counterpart, respectively. This form can be recast to a convenient material-derivative basis as $\Qdezeq = \Pdeeq\uz$ and thus, 
\begin{align}
	\left(\Pdeeq\parttone + \Qdezeq\partzone\right) = \Pdeeq\Dtuone.
\end{align}
This trick enables us to reuse the chain rule to effectively rewrite $\Pdeeq\Dtuone\Bigl\{\opHatFP,1/\opHatFP\Bigr\}$ where we note that $\opHatFP = \opHatFP\bigl(\bmu^n, \bmveta, \vT^n(T)\bigr)$ is a function of velocity polynomials of various even orders $n$, 
\begin{align}
	\Dtuone\opHatFP 	&= \left(\frac{\partial\opHatFP}{\partial\uz}\right)_{\vzeta,\vTsq}\Dtuone\uz\nonumber\\
	&+ \left(\frac{\partial\opHatFP}{\partial\vzeta}\right)_{\vTsq,\uz}\Dtuone\vzeta\nonumber\\
	&+ \left(\frac{\partial\opHatFP}{\partial\vTsq}\right)_{\uz,\vzeta}\Dtuone\vTsq.
\end{align}
Therein, we in addition to \eqref{eq:Dt_u} can identify the following material derivatives derived previously by reformulating $\vTsq,\vzeta$. Firstly, we can exploit the previously derived material derivative \eqref{eq:Dt_T} that combined with the direction-independent, root-mean-square formulation of the average thermal molecular speed $\vTsq = k_B T/m$ for molecules of mass $m$ yields,
\begin{align}
	\Dtuone\vTsq 	&= \frac{k_B}{m} \Dtuone T \nonumber\\
	&= -\frac{k_B T}{m\rho\cv}\derpTv\partzone\uz. 
\end{align}
As we have not specified a particular force model in this study we rely on Newton's second law of motion $\etaz = \Fz /m$ and \eqref{eq:Dt_p} to formulate the material derivative of $\vzeta$, 
\begin{align}
	\Dtuone\vzeta 
	&= -\frac{A\subz\rho\cssq}{\gamma m}\partzone\uz,
\end{align}
where we have further introduced the unit area $A\subz$ as a consequence of converting the force into a pressure $p = \Fz/A\subz$, such that $A\subz$ in a simulation scenario would be the grid-cell cross-sectional area. The non-ideal speed of sound $\cssq$ was  previously found in \eqref{eq:cssq}. Consequently, we obtain the explicit result for,
\begin{align}
	&\Pdeeq\Dtuone\opHatFP 	= -\Pdeeq\Biggl\{\biggl[\left(\partial_{\vzeta}\opHatFP\right)\frac{A\subz\rho\cssq}{\gamma m}\nonumber\\ 
	&\quad + \left(\partial_{\vTsq}\opHatFP\right)\frac{k_B T}{m\rho\cv}\derpTv \biggr]\bigl(\partz\uz\bigr)\nonumber\\ 
	&\quad + \left(\partial_{\uz}\opHatFP\right)\frac{1}{\rho}\bigl(\partd p \bigr)\ddz \Biggr\},
\end{align}
and similarly for the $1/\opHatFP$ destructive components with $\muFPBmns$, where we in both cases recombine $\partzone$ according to our previously defined perturbation series.

To obtain the Fourier equation we consider \eqref{eq:dttwo_T} wherein we substitute with \eqref{eq:dttwo_u}. The shear velocities arise from the derivatives of the moments,
\begin{align}
	\parttone\qdeq 	+ \parteone\Rdeeq 	= 2 \left(p - \rho\cssq\right)\partzone\uz\ud\nonumber\\ 
	+ 2p\ue\left(\parteone\ud + \partdone\ue\right) + 2p\partdone h, 
\end{align}
where $h = e + p/\rho$ is the enthalpy. As noted by \citet{Reyhanian2020, Reyhanian2021a} the resulting heat flux would be $\bmq = -\mu\nabla h$ and consequently the $g_i$-kinetic equation needs to be augmented with an energy correction term $M_0 = 2\parta\bigl(-\mu \parta h + k\parta T\bigr)$ to recover the correct Fourier law. For real gasses $\parta h = \specificcp \parta T + v\bigl(1-\beta\bigr)\parta p$ with $\beta = \rho\bigl(\partial v/\partial T\bigr)_p$ being the thermal-expansion coefficient. After applying the product rule to the shear-velocity terms, the substituted result can be recombined with the first-order perturbation contribution \eqref{eq:Dt_T} confirming the conservation of the Fourier equation, 
\begin{align}
	\rho\cv\Dtu T = -\bigl(\taude + \taudetld\bigr)\partd\ue - T\derpTv\partd\ud - \partd\qd,
\end{align}
where the normal and perturbed shear stresses organize as expected in the same formulation of the tensors (\ref{eq:taude},\ref{eq:taudetld}).

In summary, we recover the compressible, Navier-Stokes-Fourier hierarchy of equations, 
\begin{gather}
	\label{eq:NSeq_continuity}
	\Dtu\rho 		= -\rho\nabla\cdot\bmu,\\
	\label{eq:NSeq_momentum}
	\rho\Dtu\bmu 	= -\nabla p - \nabla\cdot(\tau + \tautld),\\
	\rho\cv\Dtu T 	= -(\tau + \tautld) : \nabla \bmu - T\derpTv\nabla\cdot \bmu - \nabla\cdot \bmq,
\end{gather}
in which the Fourier heat flux is defined as $\bmq = -k\nabla T$. To enable simulations we superimpose the presented viscosities into effective shear and volume viscosities, 
\begin{align}
	\mu 	&= \muB + \muFPBpls + \muFPBmns\nonumber\\
	\label{eq:mu_eff}
	&= 	\left[\left(\frac{1}{\omgBGK} - 1\right) + \left(\frac{\omgFP}{2\omgBGK}\right) - \left(\frac{1-\omgBGK/2}{\omgFP}\right)\right]p\dt,
\end{align}
\begin{align}
	\zeta 	&= \zB + \zFPBpls + \zFPBmns\nonumber\\
	\label{eq:zeta_eff}
	&= 	\left[\left(\frac{1}{\omgBGK} - 1\right) + \left(\frac{\omgFP}{2\omgBGK}\right) - \left(\frac{1-\omgBGK/2}{\omgFP}\right)\right]\nonumber\\
	&\qquad\times\left(\frac{D + 2}{D} - \frac{\rho\cssq}{p}\right)p\dt,
\end{align}
where known initial values of $\mu,\zeta$ can be defined and subsequently used to tune the time-step size $\dt$, and the kinetic relaxation frequencies $\omgBGK,\omgFP(\gamma)$, which in turn sets the thermal noise intensity governed by the Langevin equation. This result along with the kinetic equations \eqref{eq:LFPBEs} constitute the foundation of our numerical method.

\section{Fluctuation-dissipation theorem}
As a preliminary theoretical verification of the derived model, we present the fluctuation-dissipation theorems (FDTs) for the stochastic-stress tensor $\tautld$ \eqref{eq:taudetld} and compare it to the analogous FDT of the stress tensor used in Landau-Liftshitz fluctuating-hydrodynamics theory in real space and time \cite{Gallo2021,Gross2010a} namely, 
\begin{widetext}
	\begin{gather}
            \label{eq:FDT_tautld_zero}
            \bigl\langle \tauabtld(\bmx,t) \bigr\rangle = 0, \\
		\label{eq:FDT_tautld}
		\bigl\langle \tauabtld(\bmx,t)\taudetld(\bmx',t') \bigr\rangle = 2k_B T \left[\mu\left(\deltaag\dbd + \dad\dbg - \frac{2}{D}\dab\dgd\right) + \zeta\dab\dgd\right]\delta(\bmx - \bmx')\delta(t-t'),
	\end{gather}
\end{widetext}
where the shear and volume viscosities correspond to $\muB,\zB$ in \eqref{eq:taude} originating from the BGK model and the variance of the stochastic stresses is related to the hydrodynamic fluctuation energy $k_BT$. Here $\langle\cdot\rangle$ is interpreted as the steady-state average over the phase-space computed for a generic variable $\varphi$ as \cite{Gallo2022} 
\begin{align}
    \bigl\langle\varphi(\bmx,t)\bigr\rangle = N_t^{-1}N_{\bmx}^{-1}\sum_{i=1}^{N_t}\sum_{j=1}^{N_{\bmx}}\varphi_{ij},
\end{align} 
for $N_t$ time steps and $N_{\bmx}$ lattice nodes. 
We proceed with studying the behavior of a perturbed quiescent homogeneous equilibrium state at $\rho\supereq$ and a zero macroscopic velocity $\bmu = 0\suba$. If we consider only long-wavelength deviations from the equilibrium we can simplify the non-linear NSFEs into their corresponding linear forms where the macroscopic observables are linearly approximated as $\rho(\bmx,t) = \rho\supereq + \drho(\bmx,t)$ and $p(\bmx,t) = p\supereq + \delp(\bmx,t)$ \cite{Reichl1998}. By substitution into (\ref{eq:NSeq_continuity},\ref{eq:NSeq_momentum}) we obtain the linearized stochastic hydrodynamic equations, 
\begin{gather}
	\label{eq:NSE_continuity_lin}
	\partt\drho 				= -\rho\supereq\nabla\cdot \bmu,\\
	\label{eq:NSE_momentum_lin}
	\rho\supereq \partt\bmu 	= -\cssq\nabla \drho - \nabla\cdot(\tau + \tautld),
\end{gather}
in which we have expressed the bulk pressure difference in terms of the speed of sound $\delp \simeq \cssq \drho$ from its fundamental definition $\cssq \equiv (\partial p /\partial\rho)_s$. At this stage we need to isolate and decouple the longitudinal and transverse components of the velocity $\bmu$ and stochastic stress tensor $\tautld$, which is facilitated by Fourier transformations of $\drho,\bmu,\tautld$. The density Fourier transform is carried out in space as 
\begin{align}
	\label{eq:drho_Fouriertrans_space}
	\drho(\bmk,t)	= \left(\frac{1}{2\pi}\right)^{3} \int d\bmx\ \drho(\bmx,t)\exp\bigl(i\bmk\cdot\bmx\bigr),
\end{align}
and likewise for $\bmu$ and $\tautld$. Thus, Fourier-transforming the continuity equation results in \cite{Kreyszig2011a,Reichl1998}, 
\begin{align}
	\partt\drho = - i\rho\supereq\bmk\cdot\bmu.
\end{align}
Therein, $\bmu = \bmu(\bmk,t)$ can be further separated into its constituent longitudinal and transverse (perpendicular to $\bmk$) modes, 
\begin{align}
	\bmu	= \ulong\bmkhat + \bmut,
\end{align}
in which the normalized wave-vector reads $\bmkhat \equiv \bmk/|\bmk|$. Given the identity $\bmk\cdot\bmu(\bmk,t)\equiv k\ulong$ \cite{Reichl1998}, where by the definition $\bmk\cdot\bmut\equiv 0$, the continuity equation finally becomes, 
\begin{align}
	\label{eq:NSE_continuity_long}
	\partt\drho - i\rho\supereq k\ulong = 0.
\end{align}
As \citet{Gross2010a} showed for \eqref{eq:NSE_momentum_lin}, the analogous wave equations for the longitudinal $\ulong$ and transverse $\bmut\equiv\bmu\cdot\bigl(I - \bmkhat\bmkhat\bigr)$ velocities, after some amendment, simplify to, 
\begin{gather}
	\label{eq:NSE_momentum_long}
	\partt^2\ulong 	= -k^2\cssq\ulong - \nul k^2\partt\ulong + \frac{ik}{\rho\supereq}\partt \tautldl,\\
	\partt\bmut 	= -\nut k^2 \bmut + \frac{ik}{\rho\supereq}\bmtautldt,
\end{gather}
in which the longitudinal and transverse stochastic stresses are $\tautldl \equiv \bmkhat \cdot \tautld \cdot \bmkhat$ and $\bmtautldt\equiv\bmkhat\cdot\tautld\cdot \bigl(I - \bmkhat\bmkhat\bigr)$, respectively, and the associated kinematic viscosities are,   
\begin{align}
	\nul 	= \left[\zB + \muB \left(2 - \frac{2}{D}\right)\right]\frac{1}{\rho\supereq},\quad \nut = \frac{\muB}{\rho\supereq}.
\end{align}
We further note that the above results hold true for homogeneous fluids without surface tension forces, as well as square-gradient fluids since the square-gradient tension term $\rho\supereq\kappa k^2$ would absorb into the speed of sound $\cssq(\bmk) = \cssq + \rho\supereq\kappa k^2$ \cite{Gross2010a,Kim1991a}. If we Fourier transform the longitudinal wave equation in time, 
we observe that with the above foundation we arrive at the same linear-response function in frequency space $\omega$ (not to be confused with the relaxation frequencies $\omgBGK,\omgFP$) as in fluctuating hydrodynamics theories, 
\begin{align}
	\ulong(\bmk,\omega) 	= \frac{\omega k}{\rho\supereq\bigl(\omega^2 - k^2\cssq - i\omega\nul k^2\bigr)}\tautldl (\bmk,\omega),
\end{align}
in which $\tautldl (\bmk,\omega)$ acts as an ``unknown'' perturbation on the longitudinal susceptibility $\chil$ such that $\ulong(\bmk,\omega)\equiv \chil(\bmk,\omega)\tautldl(\bmk,\omega)$. In Fourier space we are interested in determining the variance matrix $A\subab(\bmk)$ which correlates the white-noise FDT $\bigl\langle|\tauabtld(\bmk,\omega)|^2\bigr\rangle = A\subab(\bmk)$. To obtain this, we further follow \citeauthor{Gross2010a} who defined the equal-time correlation function from the inverse Fourier transform $\mathF^{-1}_t$ of the ensemble average of $\ulong$,
\begin{align}
	\bigl\langle|\ulong(\bmk,t=0)|^2\bigr\rangle 	= \frac{\Al}{2\pi}\int d\omega\ |\chil(\bmk,\omega)|^2,
\end{align} 
in which $\Al$ is the longitudinal component of $A\subab$, and $|\cdot|^2$ the squared norm \cite{Lulli2024a}. Then, an explicit relation can be obtained by isolating $\Al$ with contour integration of $\int d\omega\ |\chil(\omega)|^2 = \pi/\bigl(\rho\subeq^2\nul\bigr)$ and equipartition of kinetic energy $\bigl\langle|\ulong(\bmk)|^2\bigr\rangle = k_B T/\rho\supereq$. As the variance components constitute the FDT, we find the result for the longitudinal stochastic stresses to be
\begin{align}
        \label{eq:FDT_tautldl}
	\bigl\langle|\tautldl(\bmk,\omega)|^2\bigr\rangle 	= 2k_B T\rho\supereq\nul,
\end{align}
and similarly for the transverse components, 
\begin{align}
        \label{eq:FDT_tautldt}
	\bigl\langle|\tautldt(\bmk,\omega)|^2\bigr\rangle 	= 2k_B T\rho\supereq\nut.
\end{align}
Finally, inverse-Fourier transforming these components back to real space and time results in the classical, uncorrelated FDT \eqref{eq:FDT_tautld} with $\mu = \muB, \zeta = \zB$, which is valid for simple, homogeneous fluids with white thermal noise and viscosities independent of $k$ \cite{Gross2010a,Landau1959a,Reichl1998}. 
The connection between $\tautld$ \eqref{eq:taudetld} and (\ref{eq:FDT_tautld_zero}, \ref{eq:FDT_tautld}, \ref{eq:FDT_tautldl}, \ref{eq:FDT_tautldt}) is facilitated theoretically by the hydrodynamic susceptibility $\chi$, and the FDTs are ultimately consistency requirements rather than mathematical identities derived from $\tautld$. As such, fluctuation statistics from numerical simulations can be compared to the expected variance matrix, and consistency is suggested if the evolution of $\tautld$ satisfies the ensemble averages (\ref{eq:FDT_tautld_zero},\ref{eq:FDT_tautld}). In the mathematical model and the computational realization of the LFPBM, the FDTs do not directly impose any constraints on the parameters in $\tautld$ but can be considered as an \textit{a-posteriori} numerical validation.

\section{Discussion}
\label{sec:Discussion}


As all the macroscopic observables are inherently driven by the LFPBE dynamics, it is important to convince ourselves that the reciprocal behavior between the FP-perturbations and the response in the observables evolves in a thermodynamically consistent manner, especially when the perturbations push the thermodynamic state far from equilibrium. A systematic approach for assessing this is provided by the celebrated Onsager relations \cite{Onsager1931,Onsager1953a}, which are used to set the thermal noise strength while obeying the fluctuation-dissipation theorem (FDT) and ensuring that the simulated populations in the LFPBEs meet the detailed balance conditions \cite{Gardiner1997}. Not only can they be applied to the Markovian Ornstein-Uhlenbeck process that constitutes the FPE, they can be used to observe the FDT applied to macroscopic observables \cite{Gallo2021}. This is particularly important in our derived framework, as we have heuristically added BGK relaxation into our kinetic equations \eqref{eq:LFPBEs} resulting in the viscous coefficients (\ref{eq:mu_eff}, \ref{eq:zeta_eff}) which do not yet provide a physics-informed procedure for setting the fluctuation intensity and dissipation independently of the BGK collisions. 

If we initially consider the stochastic stress tensor \eqref{eq:taudetld}, we can start with a comparison to formulations found in literature. In the Chapman-Enskog analysis carried out by \citet{Dunweg2007a}, the authors arrived at an analogous random stress tensor of the form, 
\begin{align}
	\label{eq:taudetld_Dunweg2007a}
	\taudetld	= -\frac{1}{1 - \gamma_s}\overline{R}\subde - \frac{1}{1 - \gamma_b}\frac{1}{D}\dde R\subzz,
\end{align}  
decomposed into traceless $(\delta\varepsilon)$ and trace parts $(\zeta\zeta)$. Therein, the fluctuations enforced by the zero-mean, Gaussian-random variables $R\subde$ with traceless components denoted $\overline{R}\subde$ are attenuated by the shear $(\gamma_s)$ and bulk-stress $(\gamma_b)$ dissipation parameters. These dissipation parameters are employed by the authors in a linear relaxation Monte-Carlo model, which by proxy serves an exact solution to each of the Langevin equations for the hydrodynamic modes. The random sampling performed to compute the components of \eqref{eq:taudetld_Dunweg2007a} is clearly different to our method which relies on continuously evolving $f_i,g_i(\bmvi,\bmx,t)$ and by proxy \eqref{eq:taudetld}. We would consequently expect that our stochastic stress tensor similarly is Gaussian with a zero mean such that $\langle\tautld\rangle\equiv 0$ \eqref{eq:FDT_tautld_zero}. Whether this applies in phase-space $(\bmx,t)$ considering the formalism of $\tautld$ \eqref{eq:taudetld} is not immediately clear, and will require further analysis of the fluctuation statistics from prospective simulations. We reported the FDTs for \eqref{eq:taudetld} in the previous section and showed that it adheres to the classical result of fluctuating hydrodynamics theory (\ref{eq:FDT_tautld_zero},\ref{eq:FDT_tautld}). It informs us about the behavior of thermal fluctuations in a weakly noisy environment in which wavelengths are so large that they can be rightly approximated by linearized wave equations. In our analysis, we consider the perturbation (i.e. $\tautld$) as a ``black-box'' quantity which balances these equations, and we effectively do not analyze the immediate response of varying the parameter space of $\tautld$. To complete the analysis, we will in a future work derive the covariance matrix for density, momentum, total energy, and heat flux with the purpose of realizing the computation of equilibration ratios. In turn, these would indicate if the pertinent degrees of freedom in our model are consistently thermalized.

It must be emphasized that the FPE is a deterministic diffusion equation and it is not expected that the LFPBE in its current form will be inherently able to create \textit{discrete, stochastic} noise from a uniform initial condition. Nevertheless, we do not rule out that the LFPBEs with their coarse-graining of the fluctuation information from the Langevin level will be unable to (at least partially) sustain fluctuations from an initial condition with randomly distributed variables solely via the underlying, continuous dynamics. Whether such coarse-grained information resides in the FPE and produces realistic noise spectra is unclear and needs to be treated with care, especially considering that coarse-graining approaches with non-ideal effects may incur incorrect fluctuation magnitudes from equal-time correlators \cite{Parsa2020a}. Nevertheless, the FPE has previously proven to accurately reproduce the turbulence-energy cascade down to the Markov-Einstein length scale, below which point the fluctuating statistics start to retain memory \cite{Friedrich2011,Peinke2019}. It is well known that the discrete and rarer events associated with smaller length and time scales can be accounted for by considering higher-order, non-vanishing $(n\ge 3)$ KM coefficients \cite{Peinke2019}. It is very attractive to determine the degree to which the FPE is accurate within the scope of the current research.

In prospective simulations the numerical solution strategy would roughly follow the steps: semi-Lagrangian advection including a predictor-corrector scheme predicts the populations $f_i,g_i(\bmx - \bmvi\dt,t - \dt)$ at $Q$ off-lattice departure points that comply with the local thermo-hydrodynamics at destination lattice nodes at $t$. The predictor-corrector does this by iteratively applying a linear gauge transformation to populations in a reconstruction stencil that are then used to reconstruct advecting populations at the departure points with an arbitrary interpolation kernel \cite{Dorschner2018}. Subsequently, boundary conditions are applied followed by FP perturbations and BGK collisions. The thermodynamics are imposed via a non-ideal, in our case, empirical VDW-EOS solved within the predictor-corrector scheme. As the VDW-EOS is the simplest cubic EOS with distinct energy minima and a mechanically metastable region between its spinodal and binodals, \textit{homogeneous} nucleation follows naturally when the internal energy is perturbed via thermal fluctuations.

From the Langevin equation, we stated that the FDT for the noise intensity for the Wiener process $W(t)$ in the molecular velocities evolves only in time according to the ensemble average in \eqref{eq:FDT_Langevin}. However, as the hydrodynamic dissipation attributed with the \add{stochastic} stresses $\tautld$ (and possibly a corresponding heat flux $\bmqtld$ yet to be derived) are correlated with the fields $\rhotilde, \bmutilde, \Etld$, the fluctuating variables would need to obey tailored fluctuation-dissipation \add{theorems}. In addition, $\mu,\zeta$ should similarly obey a fluctuation-dissipation \add{theorem} as they are functions of $\ptilde,\rhotilde$ and also have equilibrium values $\bigl\{\mu\supereq, \zeta\supereq\bigr\}\bigl(p\supereq,\rho\supereq\bigr)$.

The assumed white noise in the Wiener process underlying the Langevin and FPE equations represents an initial prototype for a simulation framework of thermal noise. The \add{frequency-}independence of the white noise spectrum is an idealization and may not apply accurately to all fluids, and especially not impure fluids. By exemplification, previous works on Brownian motion in water showed evidence for yellow noise scaling with the square-root of the frequency in its weak-noise spectrum in proximity to solid walls \cite{Jannasch2011}, effectively rendering the thermal noise non-Markovian, i.e. the acceleration of particles is inherently dependent on its past motion introducing a memory effect. Moreover, the colored-noise amplitude depended strongly on the distance to the wall. As such, extensions of our framework to impure, multi-component cavitating and boiling fluids may need to revisit the frequency spectrum of the thermodynamic noise.

We furthermore comment on the implications of thermal fluctuations on nucleation phenomena. Although thermal fluctuations are irrelevant in disordered, homogeneous fluid domains, thermal noise is very important in proximity to transition points, such as a spinodal where fluctuations can perturb the fluid into a new stable thermodynamic state \cite{Aursand2017}. Prospectively, assessing the fluctuating behavior around these transition points should be done rigorously on the basis of variables that exhibit sensitivity to the noise strength and correlate with the local metastable state. To that end, order-parameters $\psi(\bmx,t)$, such as that evolved by the Cahn-Hillard-Cook (CHC) equation $\partt \psi = \nabla\cdot \left[\mathcal{D}\nabla \bigl(\delta \mathcal{A}(\psi)/\delta \psi\bigr) + \tilde{\theta}(\bmx,t)\right]$ \cite{Puri2009}, which can give an indication of the mutual perturbation and similarity between non-homogenizing regions of a phase-transitioning fluid. The CHC equation is informed by a diffusion coefficient $\mathcal{D}$, the Helmholtz potential $\mathcal{A} = E(\psi) - TS(\psi)$, and lastly the noise term $\tilde{\theta}$ that obeys the FDT \cite{Callen1985, Succi2018} responsible for the correct equilibration towards the Maxwell-Boltzmann distribution. Usually, this order term contributes an additional surface-tension force $-\psi\nabla\mu_{\mathcal{A}}$ to the RHS of \eqref{eq:NSeq_momentum}, where $\mu_{\mathcal{A}}(\bmx,t) = \delta \mathcal{A}(\psi)/\delta\psi$ is the chemical potential. Additionally, the order parameter contributes a noisy-momentum flux $\nabla\cdot\tautld^{(\psi)}$ that also obeys the FDT. Evolving an equation such as the CHC may provide critical insight into metastable effects interacting with local thermodynamics and surface tension forces at interfaces. We have not considered surface-tension forces in our current single-component implementation of the FPBE, but it naturally becomes relevant in \add{multi-phase, multi-component scenarios. The CHC equation and surface-tension forces could effectively be modeled by a phase-field method, a square-gradient free energy-functional as in FLBMs \cite{Gross2010a}, or the exact-difference method applied to the Korteweg-stress tensor \cite{Kupershtokh2004a}.}

The entropy generation associated with irreversibilities from thermal fluctuations can be exactly achieved by further mathematical analysis of the FPE \cite{Lucia2014}. Ultimately, this contribution to the system entropy may be non-trivial in phase-transitioning fluids in which perturbations from noise and metastability, and therefore the characteristic relaxation rates for the phase transitions, are fundamentally dictated by the entropy-extremum principle \cite{Callen1985}. This entropy generation in turn predicts the Helmholtz and Gibbs free-energies such that there is a coupling between the noise, thermodynamics, and disorder in phase transitions. The free energy further enables predicting the degree of metastability under the saturation curve \cite{Gallo2022}.

\section{Conclusion}
\label{sec:conclusion}

To enable prospective simulation of the diffusive effects from thermal noise of Langevin particles in thermal flows, we presented a \add{two-population} lattice-Fokker-Planck-Boltzmann model of the phase-space continuous Fokker-Planck-Boltzmann equation based on Hermite quadrature and the conventional $DmQn$ lattice models. As reported in the literature, the lattice-Fokker-Planck equation usually suffers from numerical instability at lower \add{Reynolds and} Mach numbers than the Bhatnagar-Gross-Krook operator. \add{On the microscale the hydrodynamic velocity $\bmu$ is limited by the thermal velocities with the mean value $\vTsq \equiv k_B T/m$. In scenarios where $\mathO(\vTsq) \sim \mathO(\cssq)$ FLBM simulations are constrained by the threshold $\cssq \ll 1/3$ similarly to deterministic LB models applied on the macroscale \cite{Adhikari2005a}.} Thus, to mitigate sources of instability \add{in the LFPBM} in highly disequilibrated, compressible and thermal flow scenarios, we have employed the kinetic theory with discrete velocities $\bmvi = \sqrtTta\bmci + \bmu$ from Particles-on-Demand \cite{Dorschner2018}, which for the BGK operator has reported to exhibit unconditional stability. This is facilitated by the adaptivity of these velocities to the local thermo-hydrodynamics captured in the reduced pressure $\theta$ and the continuum velocity $\bmu$. Moreover, this adaptivity dovetails nicely with the introduction of the thermal-noise effects and associated non-equilibrium thermodynamic contributions of the derived lattice-Fokker-Planck-Boltzmann equations (LFPBEs) as thermal fluctuations directly influence the velocity space.

The current lattice model differentiates itself in contrast to the established FLBM by coarse-graining Langevin noise in velocity space and decomposing its effect into Kramers-Moyal drift and diffusion coefficients correlated with an underlying Langevin equation \eqref{eq:Langevin}, rather than introducing it discretely into the model variables as fundamentally stochastic processes. The drift term is fully deterministic, while the diffusion term captures the statistically continuous diffusive effects from the Langevin stochasticity in the long-wavelength limit. The thermal-noise effects are integrated into all macroscopic degrees of freedom via classical LBM hydrodynamic moments and there is no direct sampling of dynamic probability distributions during runtime.


By Chapman-Enskog multiscale analysis we showed that the two resulting lattice kinetic equations \eqref{eq:LFPBEs} with lattice-FP operators (\ref{eq:opCombi_fi}, \refeq{eq:opCombi_gi}) by proxy solve the Navier-Stokes-Fourier equations, and crucially that the dissipative effects of thermal noise induce \add{stochastic} viscous stresses $\tautld$ \eqref{eq:taudetld} where the shearing components are damped by the dynamic viscosity $\mu$ \eqref{eq:mu_eff} and the bulk terms by the volume viscosity $\zeta$ \eqref{eq:zeta_eff}. The viscosities are tuned by independently setting the relaxation frequencies $\omgFP$ and $\omgBGK$ of Fokker-Planck perturbations and BGK collisions, respectively, where $\omgFP$ directly and uniquely defines the thermal-noise strength via the friction factor $\gamma$ in the Langevin equation. \add{As an initial assessment of the correct continuum behavior, we further showed the FDT for the stochastic stress tensor reproduces the classical fluctuating-hydrodynamics balance \eqref{eq:FDT_tautld}. Our future work comprises demonstrating the correct equilibration ratios for the macroscopic variables as was done in \cite{Gross2010a} for the FLBM.}

\begin{acknowledgments}
We wish to thank C.S. Wang for his thorough introduction to the lattice-Boltzmann method, and A. Garcia and M. Sharifi for fruitful discussions. K.J.P. sincerely thanks T. Colonius for advisory support during his stay in Colonius' group, J.R. Chreim for numerous, inspiring discussions on the current research. 
Moreover, this research would not have been possible without the financial support of the Natural Sciences and Engineering Research Council of Canada (NSERC)---under Development Grant No. 519885, the American-Scandinavian Foundation, Hedorf's Foundation, and Mitacs Globalink (IT27768). K.J.P. conceptualized this research, reviewed the literature, conducted the mathematical derivations, and wrote the manuscript draft. J.R.B. provided review and editing of the written material and advisory support for the project. 
\end{acknowledgments}

\appendix

\section{Eigenfunctions of the Fokker-Planck and acceleration operators}
\label{app:Eigenfunctions}
In this section, we aim to derive the explicit eigenfunctions of the continuous FP and acceleration operators. A crucial step in discretizing the operators is to recast the entire continuous operators in convenient, functional forms that do not include any gradient terms. Reiterating, we seek two Hermite series that approximates each of the continuous operators, 
\begin{subequations}
	\begin{align}
		\label{eq:opLi_opFPi_App}
		\Bigl\{\opLi,\opFPi\Bigr\} = \wi\sumlK \frac{1}{\vTsql l!}\Bigl\{\opLaul,\opFPaul\Bigr\}\Haul(\bmvi),
	\end{align}
	where the Hermite coefficients integrate the operators acting on the population $\fvxt$, in phase space,
	\begin{align}
		\label{eq:opLaul_opLaul_App}
		\Bigl\{\opLaul,\opFPaul\Bigr\}(l) = \int d\bmv~ \Bigl\{\opHatL,\opHatFP\Bigr\}[f]\Haul(\bmv),
	\end{align}
\end{subequations}
In the following derivations we treat the \textit{hat}-operators as actions on $[f]$, but omitting the population itself. $f$ in phase-space also needs to be discretized with its Hermite coefficients $\Faul$, 
\begin{align}
	\fvxt 	= w(\bmv)\sumlK\frac{1}{\vTsql l!}\Faul(\bmx,t)\Haul(\bmv).  
\end{align}
As it turns out, one can exploit the eigenvalue property of the operators on the products $\opHatFP\bigl[w(\bmv)\Haul(\bmv)\bigr]$ identified in the Hermite series of both operators, to eliminate the derivative terms and arrive at an explicit formulation. This eigenvalue property is derived from the Gaussian,  
\begin{align}
	w(\bmv) &= \frac{1}{(2\pi\vTsq)^{\nicefrac{D}{2}}}\exp\left(-\frac{\bigl(\bmv-\bmu\bigr)^2}{2\vTsq}\right),
\end{align}
that is the common, null-space of the acceleration, FP, and BGK operators \citep{Moroni2006}. The velocity-derivative of this results in the eigenvector $-\sqrt{\theta}\ca/\vTsq$ as,
\begin{align}
	\label{eq:eigenvector}
	\partva w(\bmv) = -\frac{\sqrt{\theta}\ca}{\vTsq}w(\bmv),
\end{align}
where $\theta = p/(\rho R T_L)$ is the reduced pressure \cite{Reyhanian2020}, $\ca$ the conventional peculiar velocity for any $DmQn$ lattice model, and $\vTsq = \sqrt{RT}^2$ the squared mean thermal speed. Considering $\opFP$ excluding $f$, and omitting the implied notation $w(\bmv)$ we can expand the product, 
\begin{align}
	&\opHatFP\Bigl[w(\bmv)\Haul(\bmv)\Bigr] = \gamma \partvb\bigl(\vb + \vTsq\partvb\bigr)\Bigl[w\Haul\Bigr] \nonumber\\
	&\quad = \gamma\Bigl[\bigl(w\Haul + \vb\Haul\partvb w + \vb w\partvb\Haul\bigr) \nonumber\\
	&\quad + \bigl(\vTsq\partvb\partvb w\Haul\bigr)\Bigr].
\end{align}
Using the eigenproperty \eqref{eq:eigenvector} in the first parenthesis and rewriting the second, we obtain, 
\begin{align}
	&\opHatFP\Bigl[w(\bmv)\Haul(\bmv)\Bigr] = \gamma\Biggl[\Biggl(w\Haul - \vb\Haul\frac{\sqrtTta\cb}{\vTsq}w\nonumber\\ 
	&\quad+ \vb w\partvb\Haul\Biggr) + \vTsq\partvb\left(w\partvb\Haul + \Haul\partvb w\right)\Biggr].
\end{align}
For now we cannot manipulate the first parenthesis' terms further and we turn our attention to the second where we can use the eigenproperty as well as the chain-rule,
\begin{align}
	&(\textrm{A6}) 	= \gamma\Biggl[\left(w\Haul - \vb\Haul\frac{\sqrtTta\cb}{\vTsq}w + \vb w\partvb\Haul\right) \nonumber\\
	&\quad + \vTsq\left(\partvb\bigl\{w\partvb\Haul\bigr\} -\frac{1}{\vTsq}\partvb\bigl\{\Haul\sqrtTta\cb w\bigr\}\right)\Biggr]\nonumber\\
	&\quad = \gamma\Biggl[\left(w\Haul - \vb\Haul\frac{\sqrtTta\cb}{\vTsq}w + \vb w\partvb\Haul\right) \nonumber\\
	&\quad + \vTsq\Biggl(\bigl\{w\partvb\partvb\Haul\nonumber\\ 
	&\quad + \underbrace{\bigl(\partvb w\bigr)}_{-\frac{\sqrtTta\cb}{\vTsq}w}\bigl(\partvb\Haul\bigr)\bigr\} -\frac{1}{\vTsq}\partvb\bigl\{\Haul\sqrtTta\cb w\bigr\}\Biggr)\Biggr],
\end{align}
which yields the expanded result, 
\begin{align}
	&(\textrm{A7}) 	= \gamma\Biggl[\left(w\Haul - \vb\Haul\frac{\sqrtTta\cb}{\vTsq}w + \vb w\partvb\Haul\right) \nonumber\\
	&+ \bigl(\vTsq w\partvb\partvb\Haul - \sqrtTta\cb w\partvb\Haul\bigr) - \bigl(\partvb\Haul\sqrtTta\cb w\bigr)\Biggr].
\end{align}
Now we only need to expand the term in the third parenthesis that cf. the product rule yields, 
\begin{align}
	&(\textrm{A8}) 	= \gamma\Biggl[w\Haul - \vb\Haul\frac{\sqrtTta\cb}{\vTsq}w + \vb w\partvb\Haul\nonumber\\ 
	&\quad+ \vTsq w\partvb\partvb\Haul - \sqrtTta\cb w\partvb\Haul \nonumber\\
	&\quad - \bigl(\sqrtTta\cb w\partvb\Haul + \Haul\sqrtTta\cb\cancelto{~\eqref{eq:eigenvector}}{\partvb w}\bigr)\Biggr],
\end{align}
wherein $\sqrtTta,\cb$ are independent of $\vb$ and thus drop out with the partial derivative, such that we are left with,
\begin{align}
	&(\textrm{A9}) 	= \gamma\Biggl[w\Haul - \vb\Haul\frac{\sqrtTta\cb}{\vTsq}w + \vb w\partvb\Haul \nonumber\\
	&\quad + \vTsq w\partvb\partvb\Haul - \sqrtTta\cb w\partvb\Haul \nonumber\\
	&\quad - \left(\sqrtTta\cb w\partvb\Haul - \Haul\frac{\theta\cb\cb}{\vTsq}w\right)\Biggr].
\end{align}
In there we substitute in $\vb = \sqrtTta\cb + \ub$, such that a number of terms cancel out 
and we find the fully expanded result, 
\begin{align}
	&\opHatFP\Bigl[w(\bmv)\Haul(\bmv)\Bigr] 	= \gamma w\Biggl[1 + \ub\left(\partvb - \frac{\sqrtTta\cb}{\vTsq}\right)\nonumber\\ 
	&\quad + \vTsq\partvb\partvb - \sqrtTta\cb\partvb\Biggr]\Haul.
\end{align}
To transform the partial-derivative terms, we can exploit the Hermite relation \citep{Moroni2006a}, 
\begin{align}
	\vTsq\partvb\partvb\Haul 	= \bigl(\vb\partvb - l\bigr)\Haul,
\end{align}
so that our previous result can be rewritten to, 
\begin{align}
	(\textrm{A12})	
	&= \gamma w\left[1 - \ub\frac{\sqrtTta\cb}{\vTsq} - l + 2\ub\partvb\right]\Haul,
\end{align}
where we exploited that $\vb - \sqrtTta\cb = \ub$. Finally the remaining partial derivative can be transformed using the recurrence relation \cite{Moroni2006a, Grad1949b},
\begin{align}
	\label{eq:recurrence_rel}
	\partvb\Haul = \frac{1}{\vTsq}\Bigl[\vb\Haul - \HhermitepowLpOne\Bigr],
\end{align}
that enables us to write, 
\begin{align}
	(\textrm{A14})	
	&= \gamma w\left[1 - \ub\frac{\sqrtTta\cb}{\vTsq} - l\right]\Haul\nonumber\\ 
	&+ \gamma w \frac{2\ub}{\vTsq}\Bigl[\vb\Haul - \HhermitepowLpOne\Bigr].
\end{align}
Thereto, we can again exploit that $\vb = \sqrtTta\cb + \ub$ and factorize the $\vb\Haul$ term into the first parenthesis yielding, 
\begin{align}
	(\textrm{A16})	&= \gamma w\left[1 + \ub\frac{\vb + \ub}{\vTsq} - l\right]\Haul - \gamma w \frac{2\ub}{\vTsq}\Bigl[\Hbaul\Bigr],
\end{align}
where we further exploited that $2\vb - \sqrtTta\cb = \vb + \ub$. Thus, we find the explicit relation that can be used in the $(l\le K)$--Hermite series expansion for $\opFPi$ in \eqref{eq:opLi_opFPi_App}: 
\begin{align}
	&\opHatFP\Bigl[w(\bmv)\Haul(\bmv)\Bigr] 	= \nonumber\\
	&\uline{\gamma w\left[\left(1 + \ub\frac{\vb + \ub}{\vTsq} - l\right)\Haul - \frac{2\ub}{\vTsq}\Hbaul\right]}.
\end{align}

A similar and conveniently easier analysis can be carried out for the acceleration-operator product $\opHatL\Bigl[w(\bmv)\Haul(\bmv)\Bigr]$. Firstly, an expansion of this exploiting the linearity of the composite partial derivative renders, 
\begin{align}
	\opHatL\Bigl[w(\bmv)\Haul(\bmv)\Bigr] 	&= -\etab\partvb\Bigl[w\Haul\Bigr] \nonumber\\
	&= -\etab\Bigl[w\partvb\Haul + \Haul\partvb w\Bigr].
\end{align}
Using the eigenproperty \eqref{eq:eigenvector} and the Hermite recurrence relation \eqref{eq:recurrence_rel} this can be rewritten to, 
\begin{align}
	(\textrm{A16}) 	
	&= -\etab w \left[\frac{\Bigl(\vb - \sqrtTta\cb\Bigr)\Haul}{\vTsq} - \frac{\Hbaul}{\vTsq}\right].
\end{align}
Thus, the explicit eigenfunction we were seeking for the acceleration operator reads, 
\begin{align}
	\opHatL\Bigl[w(\bmv)\Haul(\bmv)\Bigr] 	= \uline{\frac{\etab w}{\vTsq}\Bigl[\Hbaul - \ub\Haul\Bigr]},
\end{align}
that we analogously can use in the expansion of $\opLi$ in \eqref{eq:opLi_opFPi_App}. In App. \ref{app:Hermite_coefficients} we use these to compute the Hermite coefficients for the series expansions.

\section{Hermite coefficients for the operators}
\label{app:Hermite_coefficients}

Now that we have computed the operator eigenfunctions $\bigl\{\opHatFP;\opHatL\bigr\}\Bigl[w(\bmv)\Haul(\bmv)\Bigr]$ we can establish the Hermite coefficients $\opFPaul, \opLaul$ by evaluating the integrals \eqref{eq:opLaul_opLaul_App}. For the FP operator coefficient as a composite function of $\fvxt$, we need to utilize the Hermite series \eqref{eq:fxvt_HermExp_trunc} for that population, such that the integral can be rewritten to account for $\Faul$,
\begin{align}
	\opFPgul 	&= \int d\bmv ~ \Hgul\Bigl[\opHatFP\circ f\Bigr] \nonumber\\
	&= \sumlK\frac{\Faul}{\vTsql l!}\int d\bmv ~ \Hgul\times\opHatFP\Bigl[w(\bmv)\Haul(\bmv)\Bigr],
\end{align}
where we also relocated $w\Haul$ such that the previous result in App. \ref{app:Eigenfunctions} conveniently can be identified and exploited therein. Substituting in the eigenfunction we can expand the integral to, 
\begin{align}
	\int d\bmv ~ \Hgul\Bigl[\opHatFP\circ f\Bigr] = \sumlK\frac{\Faul}{\vTsql l!}\int d\bmv ~ \Hgul\nonumber\\
	\times \gamma w\left[\left(1 + \ub\frac{\vb + \ub}{\vTsq} - l\right)\Haul - \frac{2\ub}{\vTsq}\Hbaul\right].
\end{align}
In order to compute the integral, we can exploit the Hermite orthonormality relation \cite{Moroni2006a, Grad1949b} which reads,
\begin{align}
	\label{eq:Hermite_ortho_relation}
	\int d\bmv ~ w(\bmv)\frac{1}{\vT^{l+m}}\Haul(\bmv)\Hbul(\bmv) 	= \dlm\daulbul,
\end{align}
where $\daulbul$ is the sum of products of $l$ Kronecker $\bm{\delta}$'s where each of them associates with one index in $\underline{\alpha}$ and one in $\underline{\beta}$ such that there are $l!$ terms in $\daulbul$ \cite{Grad1949b}. $\dlm$ simply denotes that only tensors of identical rank $l = m$ contribute to the sum, and all other combinations $l\ne m$ evaluate to zero for the integral. As we will have to apply \eqref{eq:Hermite_ortho_relation} twice in (B2), once one each of the $\Haul, \Hbaul$ terms, we treat them separately as follows, and combine their results after. The $l$-rank term becomes, 
\begin{align}
	&\summlK\frac{\Faul}{l!}\times\gamma \left[1 + \ub\frac{\vb + \ub}{\vTsq} - l\right]\int d\bmv ~ w\frac{1}{\vTsql}\Hgul\Haul\nonumber\\ 
	&\quad = \frac{\Faul\gamma}{l!} \left[1 + \ub\frac{\vb + \ub}{\vTsq} - l\right]\dgulaul ,
\end{align}
where we can observe that $1/\vT^{l + m} = 1/\vTsql$, as $m = l$, and that only $\Faul$ associates with the indices in $\dgulaul$. Thus, we get the discrete sum, 
\begin{align}
	\textrm{(B4)} 	= \gamma \left[1 + \ub\frac{\vb + \ub}{\vTsq} - l\right]\F^{(l)}_{\gamma_1,\dots,\gamma_l},
\end{align}
where the $l!$ sums of $\Faul$ cancelled out with the $1/l!$ prefix in (B4). Now we repeat the analysis for the second part of (B2) with $-2\ub\Hbaul/\vTsq$. In this case, we note that only the $l = m-1$ ranks of $\Hbaul$ yield a non-zero contribution to the integral. Consequently, we seek to compute the integral, 
\begin{align}
	&\summK\frac{\Faul}{\vTsql l!}\int d\bmv ~ \Hgul\times \gamma w\left[- \frac{2\ub}{\vTsq}\Hbaul\right]\nonumber\\
	&= -\summK	\frac{2\gamma\ub\Faul}{\vT^{2l + 2} l!}\int d\bmv ~ w \Hgul\Hbaul\\
	&= -\sum_{\substack{l=m-1 \\ =0}}^{K} 	\frac{2\gamma\ub\F_{\aul}^{(m-1)}}{(m - 1) !}\int d\bmv ~ w \frac{1}{\vT^{2(m - 1) + 2}} \Hgul\Hherm_{\baul}^{(m)}.
\end{align}
Therein, we note that $\Hgul, \Hherm_{\baul}^{(m)}$ are of the same ranks, and that indeed $\vT^{-[2(m - 1) + 2]} = 1/\vT^{l + m}$, such that we again can apply the orthonormality relation \eqref{eq:Hermite_ortho_relation}. Upon substitution we get the result,
\begin{align}
	\textrm{(B11)} 	= -\frac{2\gamma}{(m - 1)!}\ub\F_{\aul}^{(m-1)} \delta_{\gul,\baul}^{(m-1)}, 
\end{align}
where both $\ub, \F_{\aul}^{(m-1)}$ associate with the indices in $\delta_{\gul,\baul}^{(m-1)}$. Ultimately, we compute the discrete sum as, 
\begin{align}
	\textrm{(B12)} 	= -2\gamma\Bigl[u_{\gamma_1}\F_{\gamma_2,\dots,\gamma_m}^{(m-1)} + \cdots + u_{\gamma_m}\F_{\gamma_1,\dots,\gamma_{m-1}}^{(m-1)}\Bigr].
\end{align}
Combining the two results we get the coefficients, 
\begin{align}
	\label{eq:opFPgul_app}
	&\opFPgul(l)		= \uline{\gamma \left[1 + \ub\frac{\vb + \ub}{\vTsq} - l\right]\Fgul}\nonumber\\ 
	&\quad \uline{- 2\gamma\Bigl[u_{\gamma_1}\F_{\gamma_2,\dots,\gamma_m}^{(m-1)} + \cdots + u_{\gamma_m}\F_{\gamma_1,\dots,\gamma_{m-1}}^{(m-1)}\Bigr]}.
\end{align}
We repeat the procedure for the acceleration operator now, to which end we seek to compute the coefficients, 
\begin{align}
	\opLgul 	&= \int d\bmv ~ \Hgul\Bigl[\opHatL\circ f\Bigr]\nonumber\\
	&= \sumlK\frac{\Faul}{\vTsql l!}\int d\bmv ~ \Hgul\times\opHatL\Bigl[w(\bmv)\Haul(\bmv)\Bigr]\\
	&= \sumlK\frac{\Faul}{\vTsql l!}\int d\bmv ~ \Hgul\times \frac{\etab w}{\vTsq}\Bigl[\Hbaul - \ub\Haul\Bigr]\\
	&= \sumlK\frac{\Faul\etab}{l!}\int d\bmv ~ w\frac{1}{\vT^{2l + 2}}\Hgul\Bigl[\Hbaul - \ub\Haul\Bigr]\\
	&= \sumlK\frac{\Faul\etab}{l!}\Biggl[\int d\bmv ~ w\frac{1}{\vT^{2l + 2}}\Hgul\Hbaul\nonumber\\ 
	&\quad - \frac{\ub}{\vTsq}\int d\bmv ~ w\frac{1}{\vT^{2l}}\Hgul\Haul\Biggr],
\end{align}
where we in the last line group the terms in a convenient for aligned with the orthonormality relations. Again, letting $l \leftarrow m -1$ and then applying \eqref{eq:Hermite_ortho_relation}, the first integral becomes: 
\begin{align}
	&\sumlK\frac{\Faul\etab}{l!}\left[\int d\bmv ~ w\frac{1}{\vT^{2l + 2}}\Hgul\Hbaul\right]\nonumber\\ 	
	&\quad= \sum_{\substack{l=m-1 \\ =0}}^{K} \frac{\F_{\aul}^{(m-1)}\etab}{(m - 1)!} \left[\int d\bmv ~ w\frac{1}{\vT^{2(m-1) + 2}}\Hgul\Hherm_{\baul}^{(m)}\right]\nonumber\\
	&\quad= \frac{1}{(m - 1)!}\F_{\aul}^{(m-1)}\etab \delta_{\gul,\baul}^{(m-1)}\nonumber\\
	&\quad= \Bigl[\eta_{\gamma_1}\F_{\gamma_2,\dots,\gamma_m}^{(m-1)} + \cdots + \eta_{\gamma_m}\F_{\gamma_1,\dots,\gamma_{m-1}}^{(m-1)}\Bigr].
\end{align}
The second integral in (B18) can be directly manipulated with \eqref{eq:Hermite_ortho_relation} such that we obtain the following: 
\begin{align}
	&\sumlK\frac{\Faul\etab}{l!}\left[-\frac{\ub}{\vTsq}\int d\bmv ~ w\frac{1}{\vT^{2l}}\Hgul\Haul\right]\nonumber\\ 
	&\quad= -\summlK\frac{\etab\ub\Faul}{\vTsq l!}\left[\int d\bmv ~ w\frac{1}{\vT^{2l}}\Hgul\Haul\right]\nonumber\\
	&\quad= -\frac{\etab\ub}{\vTsq l!} \Faul\delta_{\gul,\aul}^{(l)}\nonumber\\
	&\quad= -\frac{\etab\ub}{\vTsq}\Bigl[\F^{(l)}_{\gamma_1,\dots,\gamma_l}\Bigr].
\end{align}
Combining the two integrals yields the sought after coefficients, 
\begin{align}
	\label{eq:opLgul_app}
	&\opLgul(l) 	=\nonumber\\ 
	&\uline{\Bigl[\eta_{\gamma_1}\F_{\gamma_2,\dots,\gamma_m}^{(m-1)} + \cdots + \eta_{\gamma_m}\F_{\gamma_1,\dots,\gamma_{m-1}}^{(m-1)}\Bigr]
		-\frac{\etab\ub}{\vTsq}\Bigl[\F^{(l)}_{\gamma_1,\dots,\gamma_l}\Bigr]}, 
\end{align}
that we together with \eqref{eq:opFPgul_app} can use to compute the full Hermite series. Furthermore, instead of using the $f$ distribution we can also use the $g$ distributions given that we use $\Gaul$ instead of $\Faul$ in the coefficients.


\bibliography{main_Rev2}

\end{document}